\documentclass{aa}  

\usepackage{graphicx}
\usepackage{txfonts}
\usepackage[]{hyperref}
\usepackage{graphicx}	
\usepackage{amsmath}	
\usepackage{float}
\usepackage{color}
\usepackage{soul}
\usepackage{caption}
\usepackage{subfig}

\newcommand{\gppr}{\stackrel{>}{\scriptstyle \sim}}
\newcommand{\gappr}{\raisebox{-0.4ex}{$\gppr$}}

\newcommand{\G}{{\it Gaia }}
\newcommand{\Msun}{M$_{\odot}$}
\newcommand{\mycomment}[1]{}
\usepackage{multicol}

\begin{document}

   \title{White dwarf Random Forest classification through \emph{Gaia} spectral coefficients}

   \subtitle{}

   \author{Enrique Miguel Garc\'{\i}a-Zamora
          \inst{1}
          \and
          Santiago Torres\inst{1,2}\fnmsep\thanks{Email;
    santiago.torres@upc.edu}
          \and
          Alberto Rebassa-Mansergas\inst{1,2}
                  }

\institute{Departament de F\'\i sica, 
           Universitat Polit\`ecnica de Catalunya, 
           c/Esteve Terrades 5, 
           08860 Castelldefels, 
           Spain
           \and
           Institute for Space Studies of Catalonia, 
           c/Gran Capit\`a 2--4, 
           Edif. Nexus 104, 
           08034 Barcelona, 
           Spain}

 \date{\today}
\titlerunning{Random Forest spectral classification of \emph{Gaia} white dwarfs}
\authorrunning{Garc\'{\i}a-Zamora et al.}

\offprints{S. Torres}
 
  \abstract
   {The third data release of {\it Gaia} has provided approximately 220 million low resolution spectra. Among these, about 100\,000 correspond to white dwarfs. The magnitude of this quantity of data precludes the possibility of performing spectral analysis and type determination by human inspection. In order to tackle this issue, we explore the possibility of utilising a machine learning approach, based on a Random Forest algorithm.}
   {To analyze the viability of the Random Forest algorithm for the spectral classification of the white dwarf population within 100 pc from the Sun, based on the Hermite coefficients of {\it Gaia} spectra. }
   {We utilized the assigned spectral type from the Montreal White Dwarf Database for training and testing our Random Forest algorithm. Once validated,  our algorithm model is applied to the rest of unclassified white dwarfs within 100 pc. First, we started by classifying the two major spectral type groups of white dwarfs: hydrogen-rich (DA) and hydrogen-deficient (non-DA). Next, we explored the possibility of classifying the various spectral subtypes, including in some cases the secondary spectral types. }
   {Our Random Forest classification presented a very high recall (>80\%) for DA and DB white dwarfs, and a very high precision  (>90\%) for DB, DQ and DZ white dwarfs. As a result we have assigned a spectral type to 9\,446 previously unclassified white dwarfs: 4\,739 DAs, 76 DBs (60 of them DBAs), 4\,437 DCs, 132 DZs and 62 DQs (9 of them DQpec).}
   {Despite the low resolution of {\it Gaia} spectra, the Random Forest algorithm applied to the {\it Gaia} spectral coefficients proves to be a highly valuable tool for spectral classification.}


   \maketitle
%

\section{Introduction}
\label{intro}

White dwarfs are the remnants of stars with initial masses $\la$8-10 M$_{\odot}$ \citep[e.g.][]{Althaus2010}. They are basically composed of a degenerate core of typically half a solar mass that is surrounded by a thin partially degenerate atmospheric layer. Since nuclear reactions have practically ceased, the energy source in the deep interior of white dwarfs is primarily derived from gravothermal energy released by the ions and eventually provided by core crystallization, phase separation, and other processes such as sedimentation of minor species  (see \citealt{Isern2022} for a recent review). The heat generated in the core of the white dwarf is radiated through the atmospheric envelope. Thus, this thin layer plays a capital role in the cooling of the white dwarf. In the canonical model, the outermost layer of a white dwarf is primarily composed of helium with a mass around $10^{-2}$\,\Msun, accounting for less than 2\% of the total white dwarf mass. However, in the majority of cases (approximately 80\%), there is an additional thinner layer of hydrogen with a mass between $10^{-15}$  to $10^{-4}$\,\Msun, which overlays the helium layer.

From an observational point of view, spectroscopic analysis of white dwarf atmospheres enables the identification of atomic and molecular lines and bands. This fact has allowed a spectral classification of white dwarfs attending the presence of certain lines \citep{Sion1983}. Basically, white dwarfs are divided into those that present Balmer lines (i.e. hydrogen-rich white dwarfs, or DAs), and those that do not (generically called non-DAs). Among this last group, we may also find white dwarf spectra that exhibit absorption helium lines,  \ion{He}{I} or \ion{He}{II}, called DB and DO, respectively; carbon features, either atomic or molecular, named DQ; metallic lines such as \ion{Ca}{II} or \ion{Fe}{II}, named DZ; or very weak lines or no features at all, thus showing a continuous spectrum, named DC. This general spectral classification relates to what is referred to as the primary spectral type \citep[see Table 2 from][]{Sion1983}. However, it is common to identify lines from different elements in white dwarf spectra. For instance, we may find a DA with weaker helium lines or metallic lines additionally present, in which case these objects will be labelled as DAB or DAZ, respectively. The presence of a magnetic field or variability in the white dwarf spectrum, will add a secondary H or V, respectively to the primary spectral class. 

Spectral classification of white dwarfs is of paramount importance for the determination of their stellar parameters such as temperature, surface gravity, mass or luminosity. Moreover, our understanding of the physical evolution of the white dwarf population depends on the proper identification of their atmospheres. For instance, processes like convective mixing or convective dilution in spectral evolution \citep[e.g][]{Blouin2019,Cunningham2020}, the presence of carbon in hydrogen-deficient atmospheres as a possible explanation of the {\it Gaia} color-magnitude bifurcation \citep{Camisassa2023,Blouin2023}, the high ratio of DQ white dwarfs in the so-named Q branch \citep{Tremblay2019} or the origin of accreted material in white dwarfs \citep[e.g][]{Zuckerman2007,Farihi2010}, are a few examples where a detailed identification of white dwarf spectra is required for a proper understanding of these issues.
However, spectroscopic follow-up of white dwarfs is a time cost demanding task. A  volume complete spectroscopic sample is achieved up to 40 pc from the Sun \citep{Tremblay2020, McCleery2020, Obrien2023}, but this is not the case up to 100 pc, where the percentage of spectral labelled white dwarfs is roughly 20\%  \citep[e.g.][]{Kilic2020}.

Nevertheless, the third {\it Gaia} mission Data Release \citep{GaiaDR32023} has provided astrometric data for nearly two billion objects and mean low resolution BP and RP spectra of approximately 220 million sources \citep{DeAngeli2022}. Of these,  almost 100\,000 correspond to candidates for white dwarf objects \citep{GentilleFusillo2021}.

This enormous quantity of data prevents spectral classification by human inspection. With the recent increasing growth of large astronomical databases, other approaches based on machine learning artificial intelligence algorithms are absolutely necessary. 
These techniques are widely used nowadays in astrophysics, and particularly in the field of white dwarfs. Since the pioneering work of \citet{Torres1998} on the use of self-organizing maps for the identification of halo stars, up to the most recent ones using the Random Forest algorithm in Galactic component identification \citep{Torres2019}, or their spectral identification \citep{Echeverry2022, Montegriffo2023}, or through deep learning techniques \citep{Kong2018,Olivier2023}, all of these approaches have been proven to be reliable methods in the automatic analysis of large white dwarf databases. Additional statistical classification methods have been performed, in particular in the spectral classification of white dwarfs. For instance, in \cite{Jimenez2023} and \cite{Torres2023}, the Virtual Observatory Spectral energy distribution Analyzer tool \citep{VOSA2008} was used to conduct an automated spectral energy distribution (SED) fitting of the 100\,pc and 500\,pc {\it Gaia} white dwarf samples, respectively, to different atmospheric models. These works allowed the authors to classify the samples into DA and non-DA white dwarfs with an accuracy of over 90 per cent. 

In this work we apply a Random Forest algorithm specifically developed to classify, for the first time, the whole 100-pc white dwarf \G sample into their spectral types. Focusing the analysis to objects identified within this distance limit is capital, since it represents a nearly-complete volume-limited sample, which potentially allows to derive accurate percentages of the different spectral type classes among white dwarfs. This approach based on artificial intelligence techniques represents a clear advantage respect other approaches we performed in \citet{Jimenez2023} and \cite{Torres2023} since it does not require the use of theoretical atmospheric models. The models are subject to substantial uncertainties for temperatures below 5\,500 K, which implies that unreliable classifications result for such cool white dwarfs. Instead, the Random Forest algorithm presented here relies on previously spectral type labelled white dwarfs covering all possible values of effective temperatures. Thus, we aim to obtain as much spectroscopic information as possible from the \G spectral coefficients. This includes not only a classification of the white dwarfs into their primary spectral types, but also attempting to classify them into different subcategories.

In Section \ref{s:method} we explain the methodology applied. In Section \ref{s:valid}, the validation tests performed on a subset of white dwarfs with spectral data assigned and their results are detailed. In Section \ref{s:classi}, we apply the algorithm to the classification of white dwarfs in a 100-pc radius around the Sun. In Section \ref{s:fea} we identify the most relevant spectral coefficients used in the classification process. The performance of our Random Forest algorithm is compared in Section \ref{s:others} to other recent classification methods. Finally, in Section \ref{s:conc}, we present our conclusions.

\section{The method: Random Forest classification of {\it Gaia} spectral coefficients}
\label{s:method}

The Random Forest  \citep{Breiman2001} is a widely used machine learning algorithm. From a set of labelled data, which is used to train the algorithm, an ensemble of decision trees, which is called a Random Forest, is created. Once this ensemble has been obtained, it can be used to classify new data in the given categories. 

This algorithm has been widely used for the classification of stellar objects (see, for instance, \citealt{Li2019}, \citealt{Plewa2018} or \citealt{Dubath2011}) and, in particular, to the study of the white dwarf population: some examples already show the feasibility of using Random Forest for the identification of different Galactic white dwarf populations using \G data as input parameters \citep{Torres2019} or distinguishing between spectra of isolated white dwarfs, main sequence objects and white dwarf-main sequence binaries \citep{Echeverry2022}. Moreover,  a Random Forest algorithm has also been used for the selection of white dwarfs in the {\it Gaia} sample \citep{GaiaEDR3}. Besides, a first attempt to classify white dwarfs into DAs and non-DAs using the spectral coefficients was performed by \citet{Montegriffo2023}.

Here,  following the line of the previous works,  we aimed to apply the Random Forest algorithm for the spectral classification of the {\it Gaia} white dwarf population in a 100-pc radius around the Sun. In particular, our effort is employed in classifying the different sub-populations of the non-DA sample identified by \citealt{Jimenez2023} as well as to extend the classification to cool white dwarfs ($\la5\,500$\,K) that the previous work did not consider due to the lack of accurate atmospheric models.

The {\it Gaia} spectra have low-resolution ($\lambda / \Delta\lambda \approx100$) and cover the 3\,300-10\,500 {\AA} wavelength range (3\,300-6\,800 {\AA} by the Blue Photometer (BP) and 6\,400-10\,500 {\AA} by the Red Photometer (RP); \citealt{Carrasco2021}). One particularity of the {\it Gaia} spectra is that they are not provided as a typical series of flux values for certain wavelengths but rather as a set of 55 coefficients for each of the BP and RP spectrographs (i.e. 110 coefficients in total). These coefficients refer to the Hermite functions that act as the basis for the spectral representation \citep{Carrasco2021}. The spectra are internally calibrated in a pseudo-pixel scale, and they can also be transformed to an external calibration (i.e. flux versus wavelength representation) by using the specifically designed Python package GaiaXPy\footnote{\url{https://gaia-dpci.github.io/GaiaXPy-website/}}.

As input data for the Random Forest algorithm, we use the 110 Hermite coefficients. It was demonstrated in \citet{Montegriffo2023} that the use of the coefficients provides better performance of the classification algorithm than when other input passbands were used. Moreover, the coefficient procedure can be considered totally  appropriate, as the different white dwarf spectral types are defined by their specific spectral features, and all this information is contained in the coefficients (see, for instance, \citealt{Weiler2023} for a mathematical description applied to hydrogen lines). As a consequence, no external calibration was applied, since this process may introduce what is known as `wiggles', or oscillatory behaviour. In our data, this effect would be more prominent at both ends of the externally calibrated spectra. These `wiggles' are produced by the mathematical process used to obtain the spectra \citep{DeAngeli2022}.

\section{Training and validating the algorithm: the Montreal White Dwarf database}
\label{s:valid}

The Montreal White Dwarf Database is a virtual\footnote{\url{https://www.montrealwhitedwarfdatabase.org/}} database containing astrometric, photometric and spectroscopic data including a spectral type classification for tens of thousands of white dwarfs \citep{Dufour2017}. A total of 41\,570 white dwarfs are classified into their different spectral types, 2\,905 of them within 100 pc from the Sun. In this work, the MWDD spectroscopic white dwarf classification within 100 pc is used as the input labelled sample for the training and validation tests of our Random Forest algorithm.

For the cross-validation, we adopt the stratified $k$-fold method. It consists in dividing the whole set into $k$ folds, where $k$ is a variable number (in this work we chose $k=10$). Each fold has approximately the same number of objects, and the category ratio (the proportion of objects assigned to different spectral types) is kept as close as possible to the original set category ratio. For each fold, a Random Forest is trained with all nine remainder folds, and tested on it. The advantages over the random training-test split consists in avoiding the randomness of the subset divisions, the constancy of the subset proportions and the fact that the whole set is used for both training and testing.

As it is well established, the Random Forest performance tends to be optimal for balanced data sets \citep[e.g.][]{Breiman2001}. Consequently, the validation strategy followed consisted of keeping the classification samples as close as possible to a balanced sample. Thus, our first validation test consisted in  classifying white dwarfs into DA and non-DA spectral types. The second one, focused on those labelled as non-DA, and classified them into DB, DC, DQ and DZ types. Finally, the third one consisted of classifying the white dwarfs of an specific type into its different subtypes. For instance, DA white dwarfs are divided into DA, DAB, DAH, DAP and DAZ, and similarly for the other spectral types.

In order to create the random forests and obtain the  confusion matrices and classification metrics, the Python package scikit-learn \citep{Pedregosa2011} was used. 

\begin{figure}[ht]
\includegraphics[width=1\columnwidth,trim=25 0 20 20, clip]{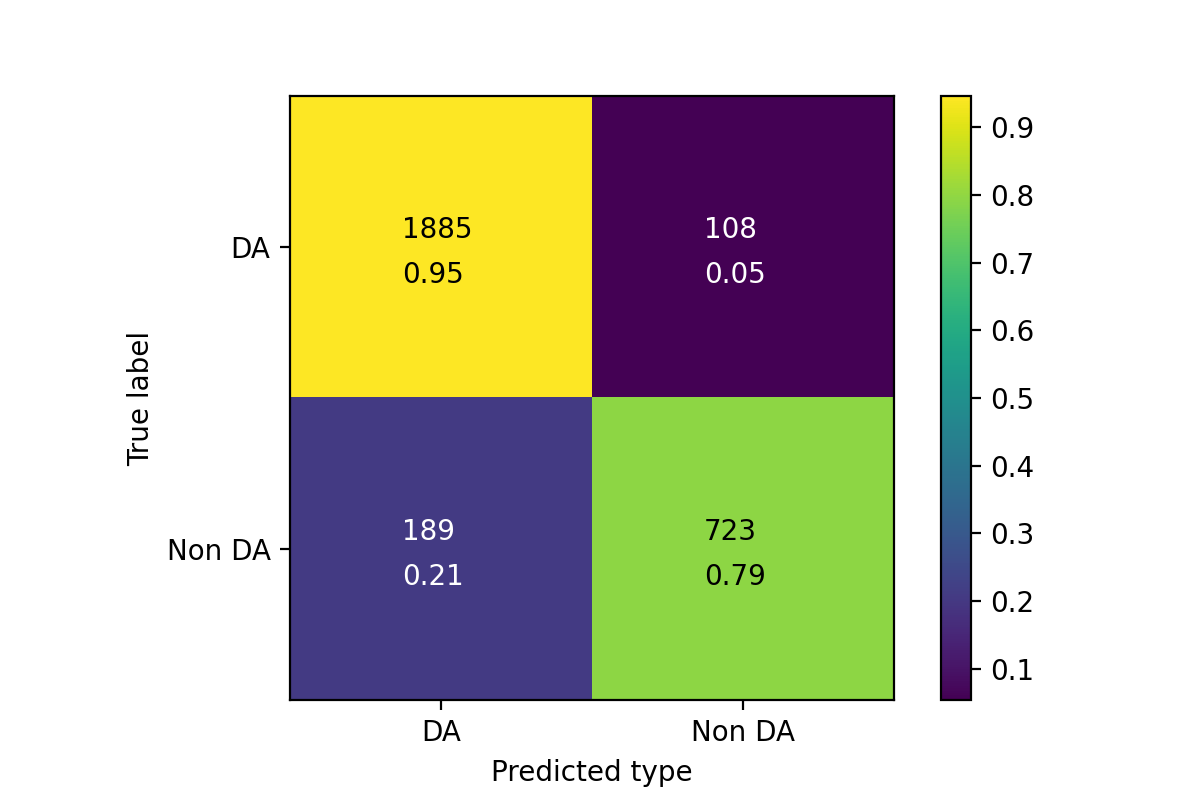}
\includegraphics[width=1\columnwidth,trim=20 0 20 20, clip]{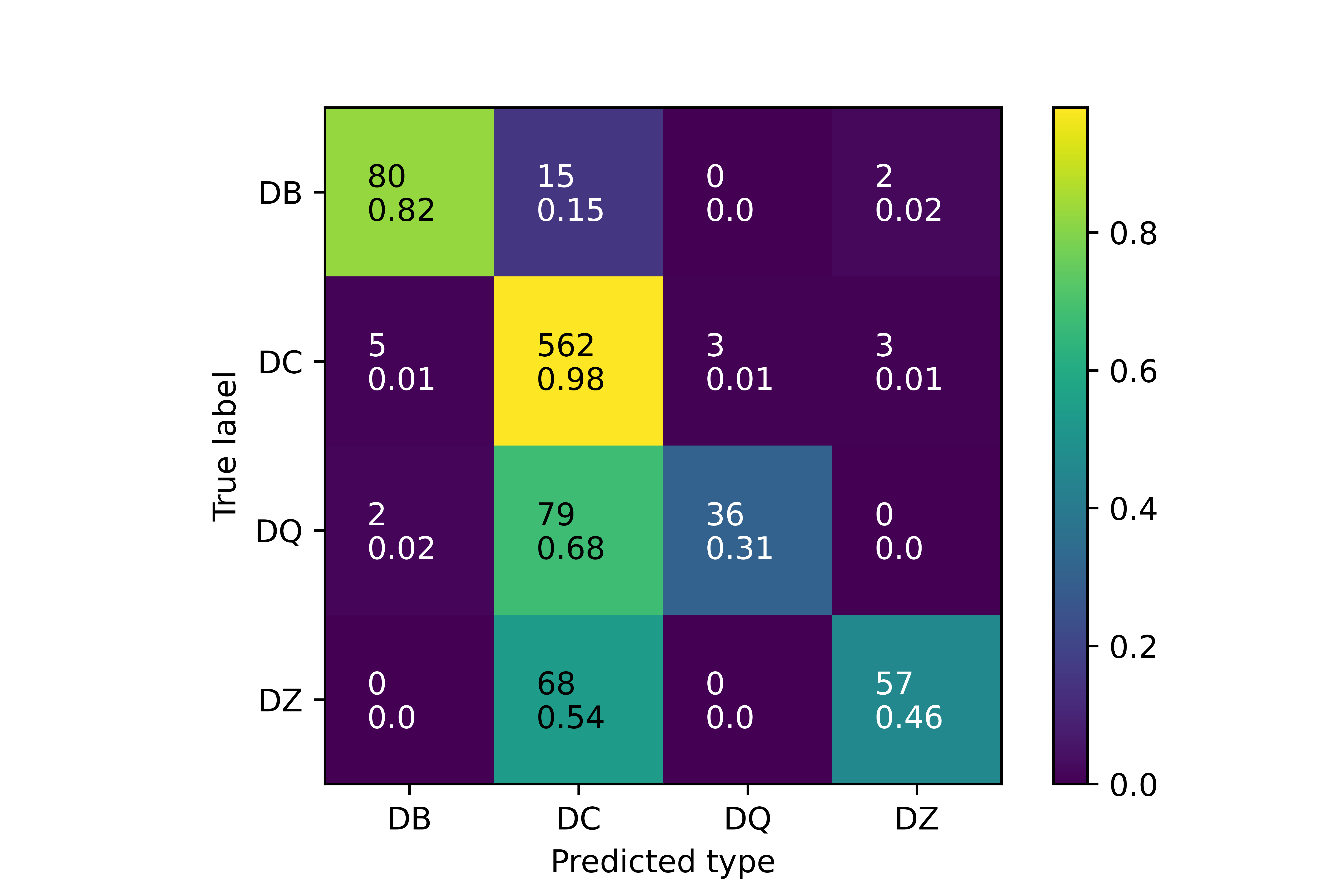}
\includegraphics[width=1\columnwidth,trim=20 0 20 20, clip]{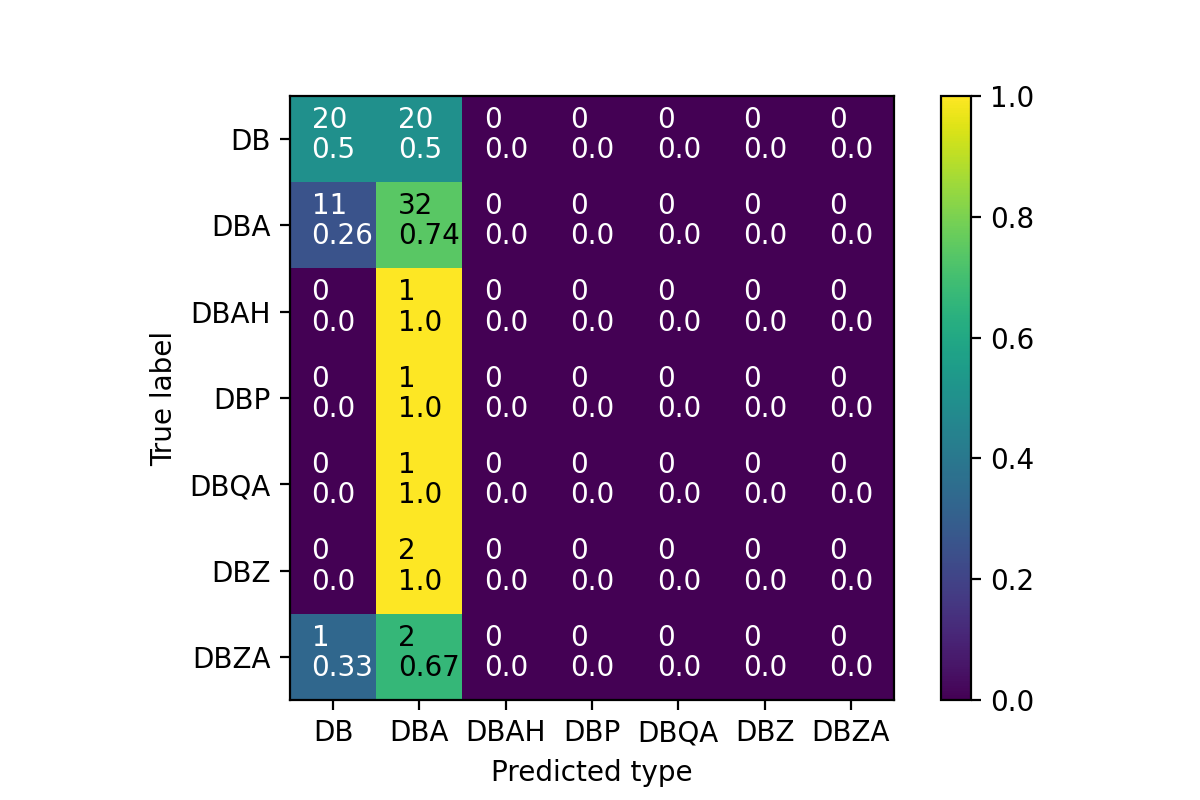}
    \caption{Confusion matrices for our validation tests: DA vs non-DA (top panel), non-DA types (middle panel) and DB subtypes (bottom panel). As true label (rows) we adopted the MWDD classification,  while the predicted label (columns) is the one resulting from our Random Forest algorithm.}
    \label{f:cmvaltests}
\end{figure}

\subsection{First validation test: classifying the {\it Gaia} population with a MWDD type into DA and non-DA types}

In the first place, we classify the whole sample of white dwarfs with labels in the MWDD within 100\,pc into DA and non-DA types (2,905 objects; 1993 as DAs and 912 as non-DAs). Although the ratio of DA and non-DA, 68.6\% and 31.4\%, respectively, is not strictly balanced, the proportion of the two groups is large enough to ensure an optimal performance of the Random Forest algorithm and avoid extreme imbalance effects. 

The resulting confusion matrix of our Random Forest model is shown in the top panel of Figure \ref{f:cmvaltests}. True labels (rows) correspond to the MWDD classification, while the label predicted by our algorithm is shown as columns. The total number of objects are indicated in each element of the matrix, and in the line below, we indicate the percentage of objects relative to a group. An ideal classification case would correspond to a diagonal matrix. The hyperparameters used for this validation test are shown in the center column of Table \ref{tab:valueshyper1} and the resulting metrics are collected in Table \ref{tab:Metrics}. For a description of the metrics, we refer the reader to Appendix A in \cite{Echeverry2022}.

The analysis of the results indicates that the performance of the Random Forest algorithm presents and excellent recall\footnote{The recall of sub-class $i$ is defined as $r_{i}={\frac{a_{ii}}{\sum_{i=1}^{n}a_{ij}}}$, where $a_{ij}$ indicate the number of objects of true class $i$ classified as class $j$.} for DA white dwarfs (95\%), and a very good recall for non-DAs (79\%). A global accuracy of 0.90 is achieved. Similar values are obtained in \cite{Jimenez2023} and \cite{Torres2023}.

\begin{table}
    \caption[]{Hyperparameters and optimal values adopted in the first two validation tests.}
    \label{tab:valueshyper1}
\begin{center}
    \begin{tabular}{lcc}
            \noalign{\smallskip}
            \hline
            \noalign{\smallskip}
                     &  \multicolumn{2}{c}{Optimal value} \\
        Hyperparameter &  Test 1 & Test 2 \\
            \noalign{\smallskip}
            \hline
            \noalign{\smallskip}
        Number of estimators & 10000 & 10000 \\

        Maximum depth & 50 & 70 \\

        Minimum samples split & 20 & 2  \\

        Minimum samples leaf & 1 & 1  \\

        Maximum features & sqrt & sqrt \\
            \noalign{\smallskip}
            \hline
    \end{tabular}
\end{center}
\end{table}

\subsection{Second validation test: classifying the {\it Gaia} non DA white dwarf population with a MWDD type into their subtypes}

In our second validation test, we classify the non-DA white dwarfs (912 objects in total) into their spectral types (DB, DC, DQ and DZ). The resulting confusion matrix is shown in the middle panel of Figure \ref{f:cmvaltests}. The hyperparameters used and the metrics obtained can be found in Tables \ref{tab:valueshyper1} and \ref{tab:Metrics}, respectively. The results obtained reveal a very good performance with an accuracy of 0.81. In particular, the algorithm presents a very good recall for DB white dwarfs (82\%), an excellent recall for DCs (98\%), and low recalls for both DQs and DZs (<50\%). Two main reasons can be identified to account for these facts. First, the low recall may be caused by the low resolution spectra inherent to {\it Gaia}. That is, not very prominent spectral lines might be unnoticed in the \G low resolution spectra, which would result in the algorithm treating them as featureless, continuous spectra characteristic of DC white dwarfs. Second, DQ and DZ classes represent 12.8\% and 13.7\%, respectively, of the non-DA population used for training. Thus, imbalanced effects, worsening the performance of the algorithm, are likely to start to manifest.

However, it must be noted that, in spite of the low recall for DQ and DZ white dwarfs, their precision\footnote{Analogously to the recall, the precision of sub-class $i$  is defined as $p_{i}={\frac{a_{ii}}{\sum_{j=1}^{n}a_{ij}}}$} (as well as for DB white dwarfs) is excellent, i.e. 92\% for the three types. False positives are almost absent. This implies that, while the algorithm does not find all DZ and DQ white dwarfs, the probability of a white dwarf belonging to the type it has been classified into is very high. This makes our algorithm highly useful for efficiently identifying white dwarfs of these spectral types within an unclassified population.

Finally, as a verification exercise, we have attempted to classify the entire sample into their primary subtypes; DA, DB, DC, DQ and DZ. An excellent recall is achieved for DAs (97 \%), DBs show a very good recall (80\%); and DQ and DZ recalls are certainly improvable (25 and 34\%, respectively). Once more, DBs, DQs and DZs show an excellent precision (89\% for DBs and 91\% for DQs and DZs). However, the scoring values are lower than the values obtained in the first two validation tests. The conclusion we extract from this result is that better results are obtained when the workflow includes a first DA/non-DA classification and a second, specific non-DA classification.

\begin{table*}[h!]
    \caption[]{Classification metrics for the first validation test in which we classify white dwarfs of the MWDD into DA and non-DA classes, and the second validation test in which non-DAs are classified into DB, DC, DQ and DZ.}
    \label{tab:Metrics}
\begin{center}
    \begin{tabular}{lcccccccc}
            \noalign{\smallskip}
            \hline
            \noalign{\smallskip}
        Metric & Validation test & DA  & non-DA & DB & DC & DQ & DZ   \\
            \noalign{\smallskip}
            \hline
            \noalign{\smallskip}
        Recall & Test 1 &  0.95 & 0.79 & - & - & - & -  \\
        Recall & Test 2 & - & - & 0.82 & 0.98 & 0.31 & 0.46  \\
            \noalign{\smallskip}
            \hline
            \noalign{\smallskip}
        Precision & Test 1 & 0.91 & 0.87 & - & - & - & -  \\
        Precision & Test 2 & - & - & 0.92 & 0.78 & 0.92 & 0.92  \\
            \noalign{\smallskip}
            \hline
            \noalign{\smallskip}
        F1 score & Test 1 & 0.93& 0.83 & - & - & - & -  \\
        F1 score & Test 2 & - & - & 0.87 & 0.87 & 0.46 & 0.61   \\
            \noalign{\smallskip}
            \hline
    \end{tabular}
\end{center}
\end{table*}

\subsection{Third validation test: classifiying the {\it Gaia} white dwarf population with a MWDD type into their secondary types}
\label{s:third}

So far, we have demonstrated that the Random Forest algorithm based on the coefficients of {\it Gaia} spectra is a feasible tool for classifying white dwarfs into their primary spectral types. Now, we explore the possibility to classify the secondary types. Several factors prevent us from being optimistic about such performance. First, {\it Gaia} low-resolution spectra appear to have limited capability to discern detailed features, such as distorted Balmer lines in DAH white dwarfs or weak lines in other atmospheres. Second, subtype classes represent in many cases clearly imbalanced samples with respect to the predominant subtype, thus worsening the performance of the algorithm.

Nevertheless, even if we expect a very low recall in the classification of the majority of spectral subtypes, if a high precision is achieved, it would imply the identification of valuable objects. Consequently, the five considered spectral types (DA, DB, DC, DZ, and DQ) were divided into different subtypes and analyzed separately.

\subsubsection{DA subtype classification}
\label{3.3.1}

The DA type was divided into pure DA, DAB, DAH, DAP and DAZ. The classification test reveals the disappointing, but not unexpected, result that only one DAH white dwarf is correctly classified as such (1\% recall). This shows that, in almost all cases, the fine magnetic splitting of the spectral lines produced by the intense magnetic field is too small to be discerned in the low resolution {\it Gaia} spectra. Additionally, one DA is missclassified as a DAH, resulting in a precision of 50\% for DAHs.

The rest of the subtypes are not recognized in the classification and are all missclassified as DAs. Low prominent lines and the spectral resolution, as well as their low number compared to the initial sample, would explain these results.

\subsubsection{DB subtype classification}

In this test, the DB type was divided into pure DB, DBA, DBAH, DBP, DBQA, DBZ and DBZA. Except for DBAs, which comprise 47\% of the DB sample, the other subtypes are only a residual part. Imbalanced effects are therefore expected. 

The confusion matrix (bottom panel of Figure \ref{f:cmvaltests}) shows that, except for the DBA type, all subtypes are misclassified. DBAH, DBP, DBZ and DBQA subtypes are incorrectly classified as DBA; and DBZA are missclassified as 33\% DB and 67\% DBA. In the DB/DBA classification, 50\% of the DB and 74\% of DBAs are correctly classified.

\subsubsection{DC subtype classification}

Considered DC subtypes include pure DC, DCP and DCQ. However, the algorithm is, as expected, unable to distinguish between these different subtypes: all objects in this subset are classified as DC. Two possible explanations are proposed for this result. First, the {\it Gaia} spectral resolution is likely too low for the subtype-defining spectral classes to be noticeable. Second, the extremely low number of DCP and DCQ objects (two for each subtype against 569 objects labelled as DC) is not enough for the algorithm to be able to properly differentiate between them.

\subsubsection{DQ subtype classification}

The results of the Random Forest applied to the DQ subtype reveals that the algorithm is only able to distinguish DQpec white dwarfs, and even then only two out of seventeen (11\% recall). The other subtypes (DQA, DQZ and DQZA), which comprise at most two objects for each, are indistinguishable from DQs.

It is also worth noting, however, that precision is also perfect for the DQpec subtype (100\%), with no false positives from other subtypes. This implies that the few DQpec stars the algorithm may find have a very high probability of belonging to this subtype.

\subsubsection{DZ subtype classification}

Regarding DZs, the Random Forest algorithm is not able to distinguish DZH and DZP subtypes from DZ white dwarfs. On the other hand, a single DZA (8\% recall) is properly classified. However, three DZs are mislabelled as DZA, negatively impacting the precision for this subtype (25\%).

\subsubsection{Spectral subtype classification summary}

From the analysis of the results of the Random Forest algorithm applied to the different spectral subtypes, we can conclude that the algorithm is mostly unable to classify secondary spectral subtypes, whether due to the numerical imbalances or the inherent low resolution of the {\it Gaia} spectra that prevents their spectral lines from being recognised by the algorithm.

A possible exception is the subtype DQpec, which, although shows a low recall, also presents a perfect precision. Its situation among DQ subtypes is similar to the situation of DQs among non-DA white dwarfs: low recall, but very high precision that might allow us to find candidates with a very high probability of actually belonging to the group is has been classified into.

Furthermore, although the DB/DBA classification may seem possible due to the good recovery of DBA white dwarfs, we must take this result cautiously. While the recall is reasonably good for DBAs (74\%), it is just 50\% for DBs. Additionally, the precision is only slightly superior to 50\% for both subtypes. Consequently, we cannot assume that a white dwarf identified as a DB or DBA has a high probability of really being one.

\section{Classifying the {\it Gaia } non-DA 100-pc white dwarf population}
\label{s:classi}

Once our Random Forest algorithm has been tested and validated, it can be applied to the unclassified {\it Gaia} 100-pc white dwarf population. A subgroup of these white dwarfs, namely those with ${\rm BR-RP}<0.86$ (equivalent to white dwarfs hotter than $\ga$5\,500\,K), has already been classified into DAs and non-DAs by \cite{Jimenez2023}. To that end, synthetic photometry of all white dwarfs was generated using their spectra and the J-PAS \citep{Benitez2014} filter system \citep{Marin2012}. These spectra were fitted using a collection of DA and DB atmospheric models, and a probability for each them belonging to the DA type was computed from the $\chi^2$ arising from the best fits. In this exercise, \cite{Jimenez2023} adopted two approaches: model fits using all \G spectral coefficients and model fits using the truncated coefficients. The former case, defined as the VOSA-GJP estimator, provided better results, with an overall accuracy of 91\%. This value is slightly higher than the accuracy we have obtained here using our Random Forest algorithm (90\%). Although both classification performances can be considered practically equivalent, we hereafter adopt the DA and non-DA VOSA-GJP classification of \cite{Jimenez2023} for all white dwarfs with ${\rm BR-RP}<0.86$.

Thus, in this section we first apply our Random Forest model to those white dwarfs classified into DA and non-DA by \cite{Jimenez2023} with the aim to obtain their spectral subtypes. Then, we expand the classification to those unclassified objects with colour ${\rm BR-RP}>0.86$, i.e. the cooler white dwarfs that we failed to identify in \cite{Jimenez2023} due to the lack of accurate atmospheric models.

\subsection{White dwarfs identified by VOSA-GJP}

In this section, we analyse the objects classified in \cite{Jimenez2023} with the aim of obtaining their spectral syb-types. In that classification, as mentioned before, white dwarfs were assigned a probability, $P_{\rm DA}$, of being DAs. Those with $P_{\rm DA}>0.5$ were classified as DAs, while those with $P_{\rm DA}<0.5$ were classified as non-DAs. A total of 5\,823 white dwarfs with ${\rm BR-RP}>0.86$ are considered in this section; 4\,157 of them classified as DAs and 1\,666 classified as non-DAs.

\subsubsection{DA white dwarfs identified by VOSA-GJP}
\label{4.1.1}

\begin{figure}[ht]
    \includegraphics[width=1.0\columnwidth,trim=10 0 30 20, clip]{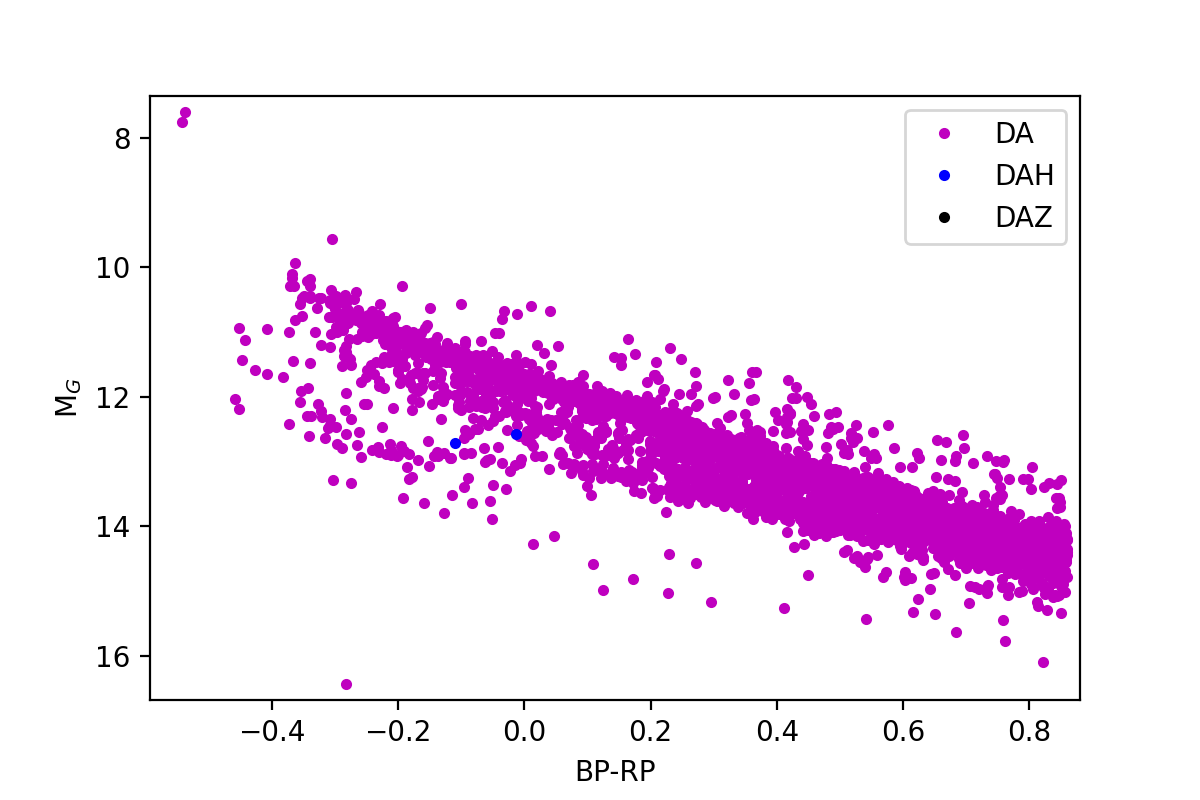}
                \caption{HR diagram of the classified {\it Gaia} DA 100-pc white dwarf population in \cite{Jimenez2023} Two DAH (blue dots) are identified by our algorithm.}
    \label{f:DA-DAH}
\end{figure}

Despite the poor performance of the algorithm in classifying secondary spectral types found in subsection \ref{3.3.1}, we attempted to find possible DAH or DAZ candidates among the group of 4\,157 white dwarfs classified as DA in  \cite{Jimenez2023}.

The result reveals that only two DAH candidates are found. Nonetheless, despite the very low number of DAHs, we consider it as a success for our algorithm, specially since the effect of magnetic fields in spectral lines is fine magnetic splitting, which is not easily noticeable in low-resolution spectra. In Figure \ref{f:DA-DAH} we show the location of these two DAH candidates in the {\it Gaia} Hertzsprung-Russell (HR) diagram.

\subsubsection{Non-DA white dwarfs identified by VOSA-GJP}
\label{4.1.2}

We analyze now the white dwarfs that have been classified as non-DAs in \cite{Jimenez2023} via adopting the VOSA-GJP estimator. In order to train the set, we once more resort to the MWDD. From the whole set, we derive a subset that mimics the conditions of the objects that will be classified (i.e. non-DA white dwarfs with ${\rm BR-RP}<0.86$). This left us with only 509 objects in the training set for the Random Forest algorithm, contrasting with 912 non-DA white dwarfs classified in the whole training set.

\begin{figure}[h]
    \includegraphics[width=0.95\columnwidth,trim=10 0 30 20, clip]{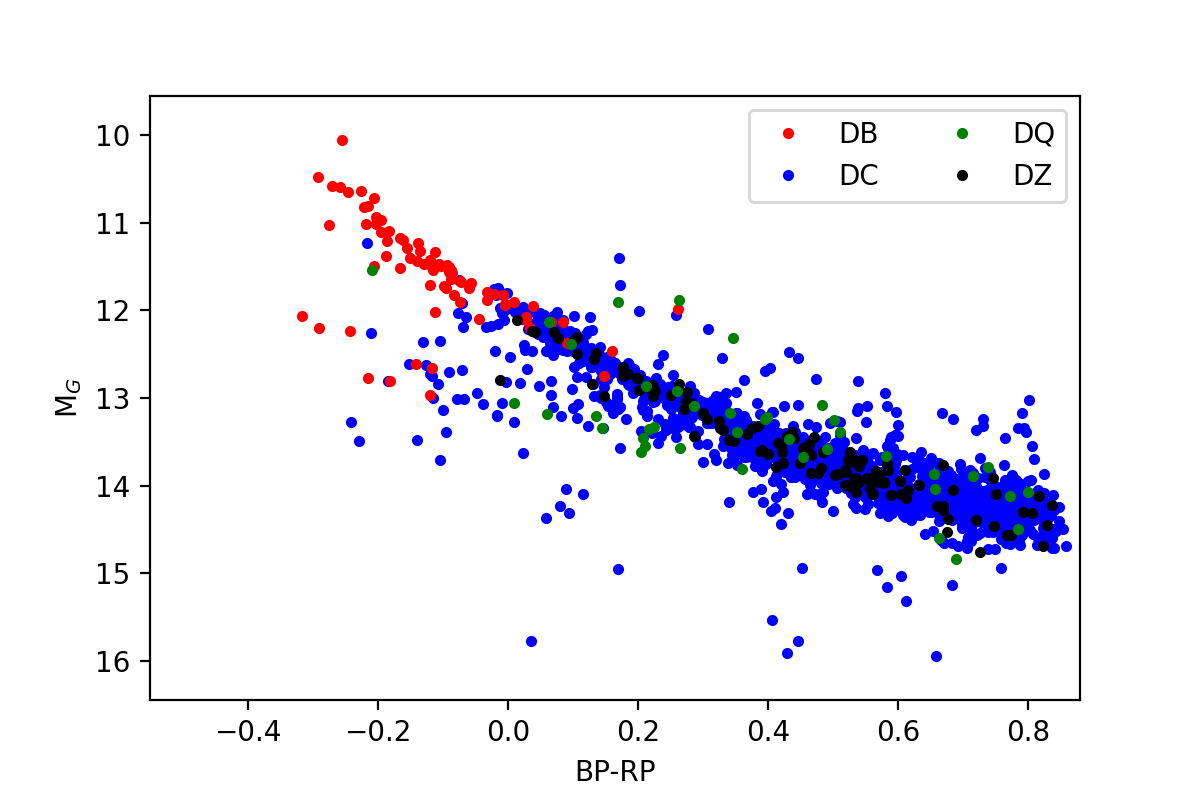}
                \caption{HR diagram of the classified {\it Gaia} non-DA white dwarf population within 100 pc in \cite{Jimenez2023}, divided into their different subtypes found in this work.}
    \label{f:nonDAHR}
\end{figure}

The classifying algorithm is then applied to the rest 1\,666 objects in the test subset. The classification yields the following results: 76 objects are identified as DBs, 1\,429 as DCs, 40 as DQs and 121 as DZs. The corresponding HR diagram of these classified objects is shown in Figure \ref{f:nonDAHR}.

The HR diagram not only serves to illustrate the composition of the classified population, but it also allows us to check for consistency with expected white dwarf characteristics. For instance, no DB white dwarfs should be found below a certain temperature ($\approx10\,000\,$K). In Figure \ref{f:nonDAHR}, DBs appear restricted to the top left, hotter region of the white dwarf sequence, while some DQ white dwarfs appear in the DQ branch. All these factors reinforce the idea that our classification is essentially correct, and no spectral types appear outside of their expected locations.

\begin{figure}[ht]
    \includegraphics[width=0.95\columnwidth,trim=10 0 30 20, clip]{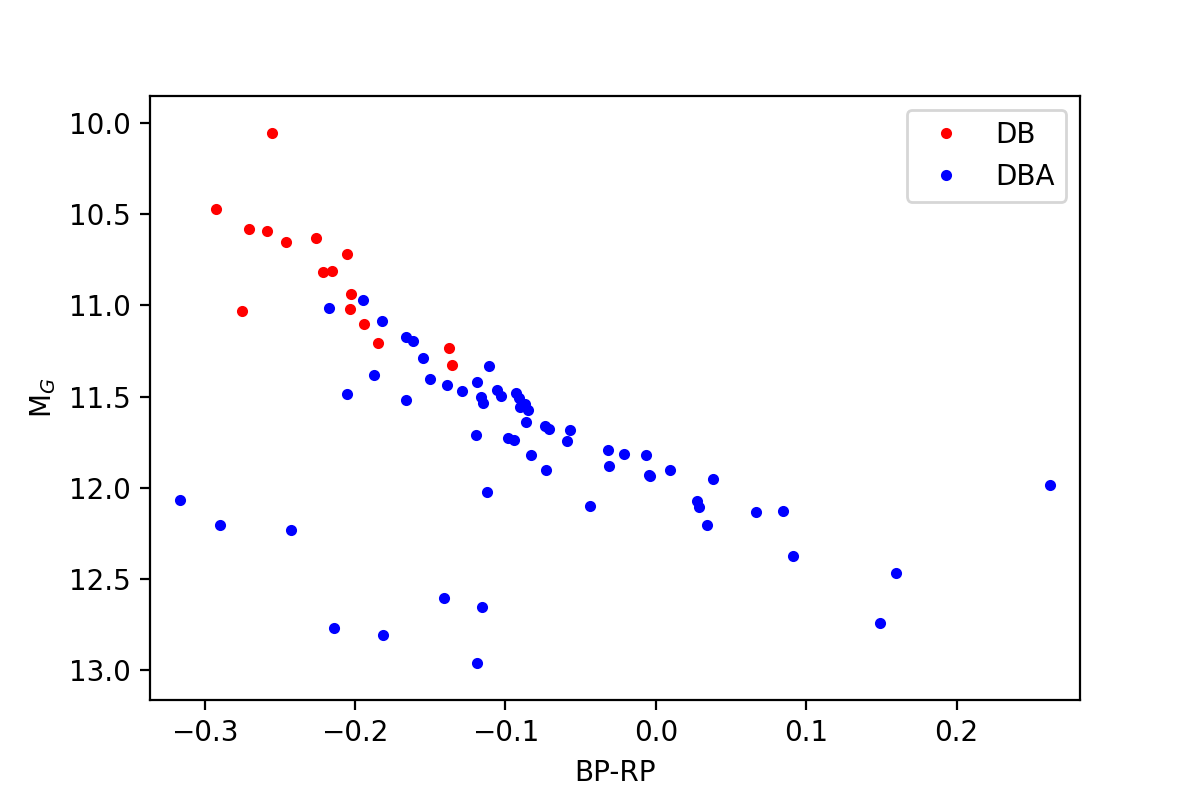}
                \caption{HR diagram of the classified pure DB and DBA white dwarfs found in this work.}
    \label{f:DBDBA}
\end{figure}

Furthermore, as seen in our third validation test (see Section \ref{s:third}), the Random Forest algorithm is able to identify (although with low recall but with high precision) secondary subtypes of DBs and DQs.  As we do not expect to find more DBs in the cooler region that remains to be analyzed, we apply our classification algorithm to the set of 76 DB white dwarfs identified so far. The Random Forest identified 16 pure DB and 60 DBA objects. No other secondary subtypes (DBAH, DBAZ, DBP, DBQA, DBZ, DBZA and DBe) were identified. In Figure \ref{f:DBDBA} we depict the HR diagram location for the identified pure DBs and DBAs. We can check that  no pure DBs are found with colors redder than ${\rm BP-RP}\gppr-0.1$ (i.e. effective temperatures cooler than $\approx$12\,000\,K), this fact being clearly indicative of the presence of hydrogen in pure-helium atmospheres due to convective mixing \citep[e.g. see][and references therein]{Bergeron2019}.

The sample of identified DQs is analyzed into its secondary types in Section \ref{s:DQsubtypes}.

\subsection{White dwarfs not identified by VOSA-GJP}

Once those white dwarfs classified  in \cite{Jimenez2023} as DA and non-DA have been further classified in their different subtypes, it was decided to explore the cold region of the HR-diagram, i.e. ${\rm BR-RP}>0.86$, which had not been analyzed in the aforementioned work. 

As described in Section \ref{s:valid}, our strategy consists in first classifying the cold white dwarf sample into DAs and non-DAs; then,  the non-DAs  are classified into DBs, DCs, DQs and DZs. Finally, we look for possible secondary spectral type candidates (although this last step will probably be impracticable, as the spectra in this region have a very low signal-to-noise ratio).  

\subsubsection{DA vs non-DA classification.}
\label{4.2.1}

The number of white dwarfs present in the 100 pc sample from \cite{Jimenez2023} and with colors ${\rm BR-RP}>0.86$ (that is, with no VOSA-GJP classification) is 3\,623 objects. To that sample we apply our Random Forest algorithm, once trained with those objects labelled in the MWDD (2\,905 objects). 
It is worth saying that the MWDD sample contains 192 DA and 293 non-DA white dwarfs with colors ${\rm BR-RP}>0.86$ and within 100 pc. This set of white dwarfs guarantees the reliability of our method, as there are enough labeled objects to train the algorithm in the HR region of interest.

\begin{figure}[t]
    \includegraphics[width=0.95\columnwidth,trim=10 0 30 20, clip]{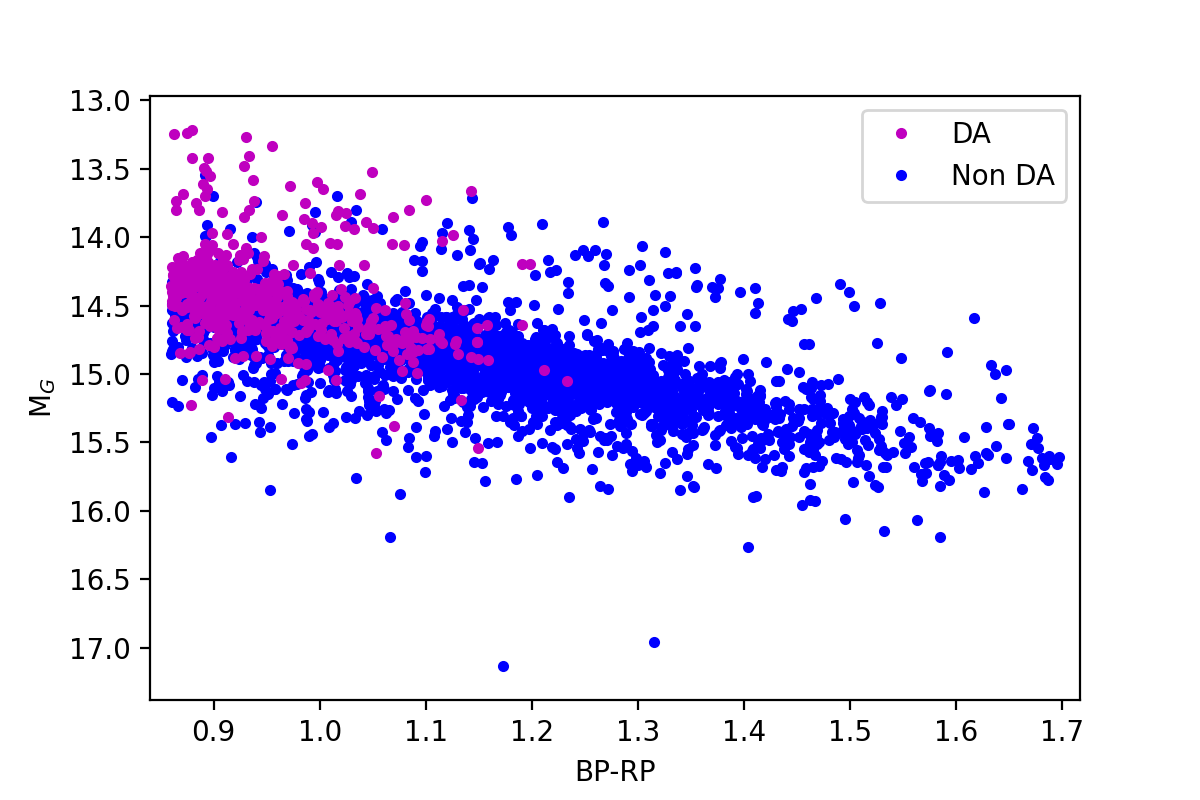}
                \caption{HR diagram of the classified {\it Gaia} DA and non-DA 100-pc white dwarf population with color BP-RP$>0.86$. }
    \label{f:HRcold}
\end{figure}

The results of applying our Random Forest to the cold sample of \cite{Jimenez2023} can be seen in the HR diagram presented in Figure \ref{f:HRcold}. The vast majority of white dwarfs (3\,041; 84\%) are classified as non-DAs (blue dots), while DAs (magenta dots) comprise only a small fraction (582; 16\%) and none of them has colour ${\rm BR-RP}>1.25$. 

This classification is consistent with the expected behaviour at  temperatures lower than $\simeq$5\,000\,K, since at this range the hydrogen in the white dwarf atmosphere remains mostly in the ground state. Thus, Balmer spectral lines would become too weak (or they simply disappear) to be detected in {\it Gaia} low resolution spectra and, consequently, the object will be classified as a featureless DC.

\subsubsection{DA secondary type classification}

As we have seen in Section \ref{4.1.1}, our Random Forest algorithm was able to find two DAHs. Thus, we applied the algorithm to the classified cold DA white dwarfs. From the 582 objects, none was classified as a DAH or DAZ; all of them were classified as DAs. This result is not entirely unexpected as it was already known that the {\it Gaia} resolution was, in almost all cases, insufficient for this purpose.

\begin{figure}[t]
    \includegraphics[width=0.95\columnwidth,trim=10 0 30 20, clip]{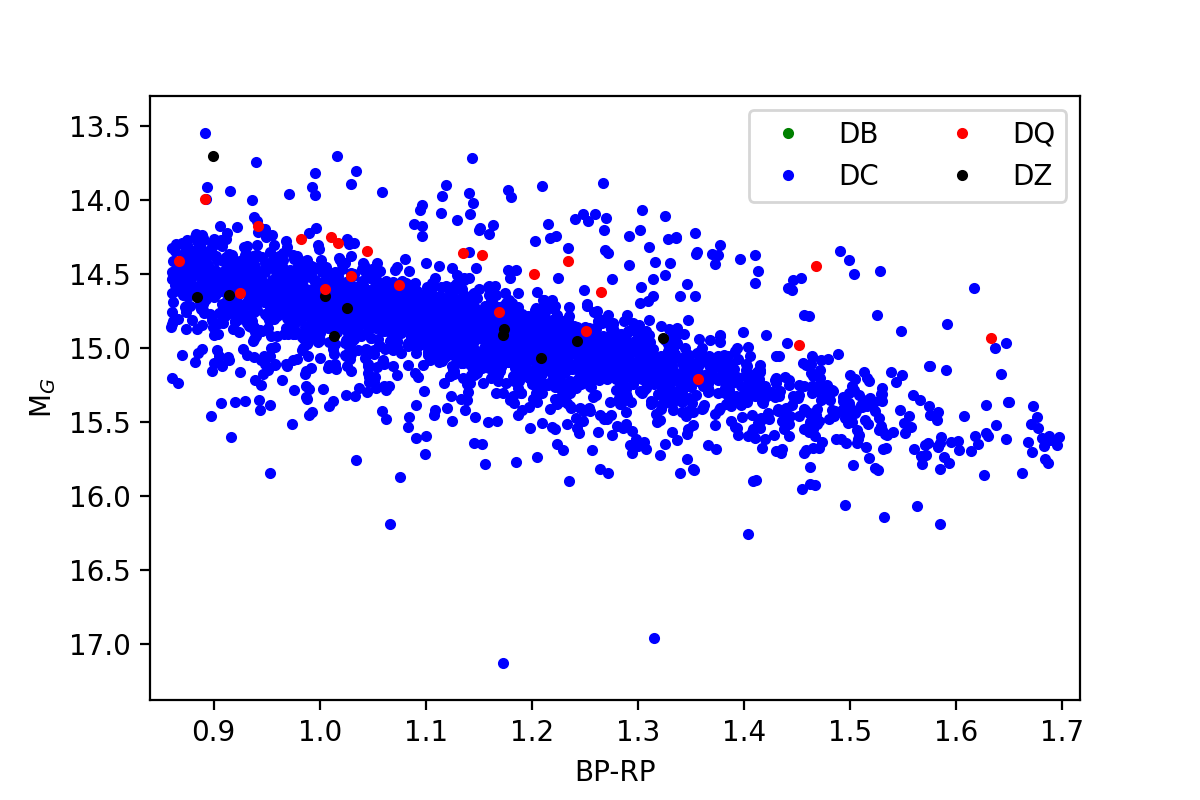}
                \caption{As Fig. \ref{f:HRcold}, but showing the classification of non-DAs into their different spectral subtypes.}
    \label{f:HRcoldnoDA}
\end{figure}

\begin{figure}[ht]
    \includegraphics[width=0.95\columnwidth,trim=10 0 30 20, clip]{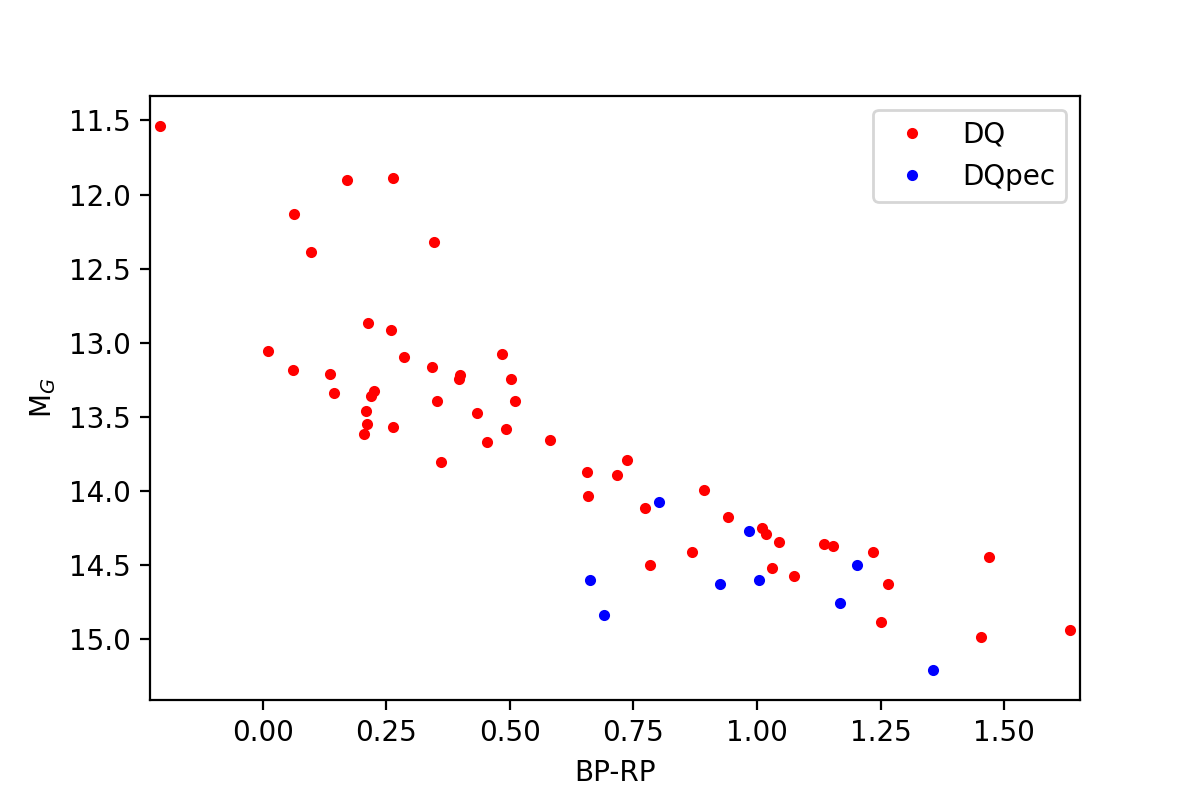}
                \caption{HR diagram of the classified DQ and DQpec white dwarfs found in this work.}
    \label{f:DQpec}
\end{figure}

\subsubsection{Non-DAs subtype classification.}

Once the identification between DAs and non-DAs has been completed, the 3\,041 found non-DAs were classified into DC, DQ and DZ categories. DBs were discarded, as none can be found at these low temperatures. As in subsection \ref{4.2.1}, the whole MWDD classified set was used as the training data. In particular, we used 248 DCs, 19 DQ and 26 DZ with colours ${\rm BR-RP}>0.86$.

The results shown in Figure \ref{f:HRcoldnoDA} reveal, as expected, that the most prominent group are DCs: 3\,008 objects representing  98.9\% of the sample. Despite the low \G resolution, and possibly low S/R, which impair the algorithm's ability to correctly identify any spectral feature at this low temperature regime, the Random Forest algorithm was able to identify 22 DQs and 11 DZs. Taking into account that only 19 DQs and 26 DZs white dwarfs with colours ${\rm BR-RP}>0.86$ form the training sample, these newly found objects represent a 115.6\% and 42.3\% increment, respectively.

\subsection{DQ secondary type classification}
\label{s:DQsubtypes}

In our analysis, we have found 62 DQ so far. Of them, 40 with colors bluer than BP-RP$=0.86$ (see Section \ref{4.1.2}) and 22 redder than that value (previous section). As demonstrated in our third validation test (see Section \ref{s:third}), the Random Forest is capable to identified certain secondary DQ spectral types, although at the expense of low recall. 

We apply our Random Forest algorithm to the 62 DQ-identified white dwarfs, with the aim of classifying them into the secondary spectral types (i.e. DQ, DQA, DQP, DQZ, DQZA, DQp, DQpec and DQpecP). The result reveals that objects are thus classified only into two groups: 53 DQ and 9 DQpec white dwarfs. Figure \ref{f:DQpec} shows the corresponding HR diagram. Those objects classified as DQpec are tipically cold, with BP-RP$\gappr0.6$, indicative that Swan bands are more easily distorted at low temperatures \citep[e.g][]{Blouin2019b}.

\begin{figure}[h]
    \includegraphics[width=1.\columnwidth,trim=10 0 10 20, clip]{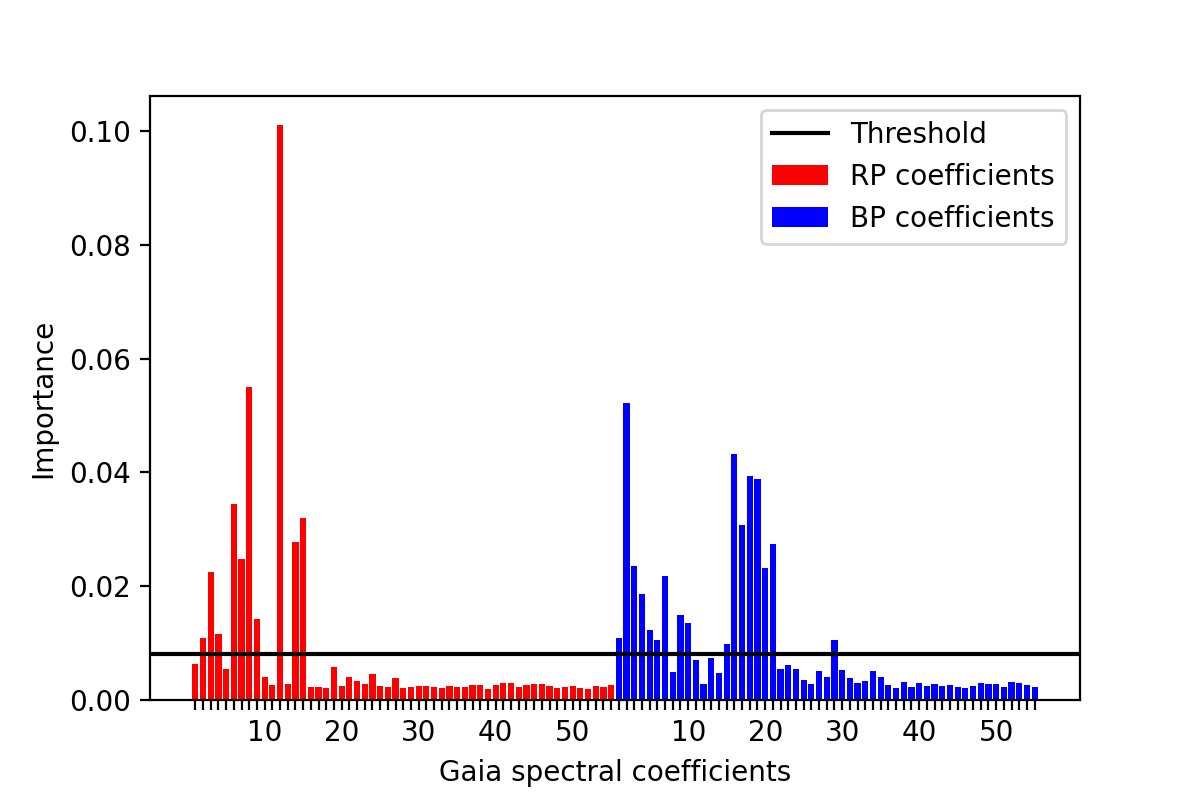}
    \includegraphics[width=1.\columnwidth,trim=10 0 10 20, clip]{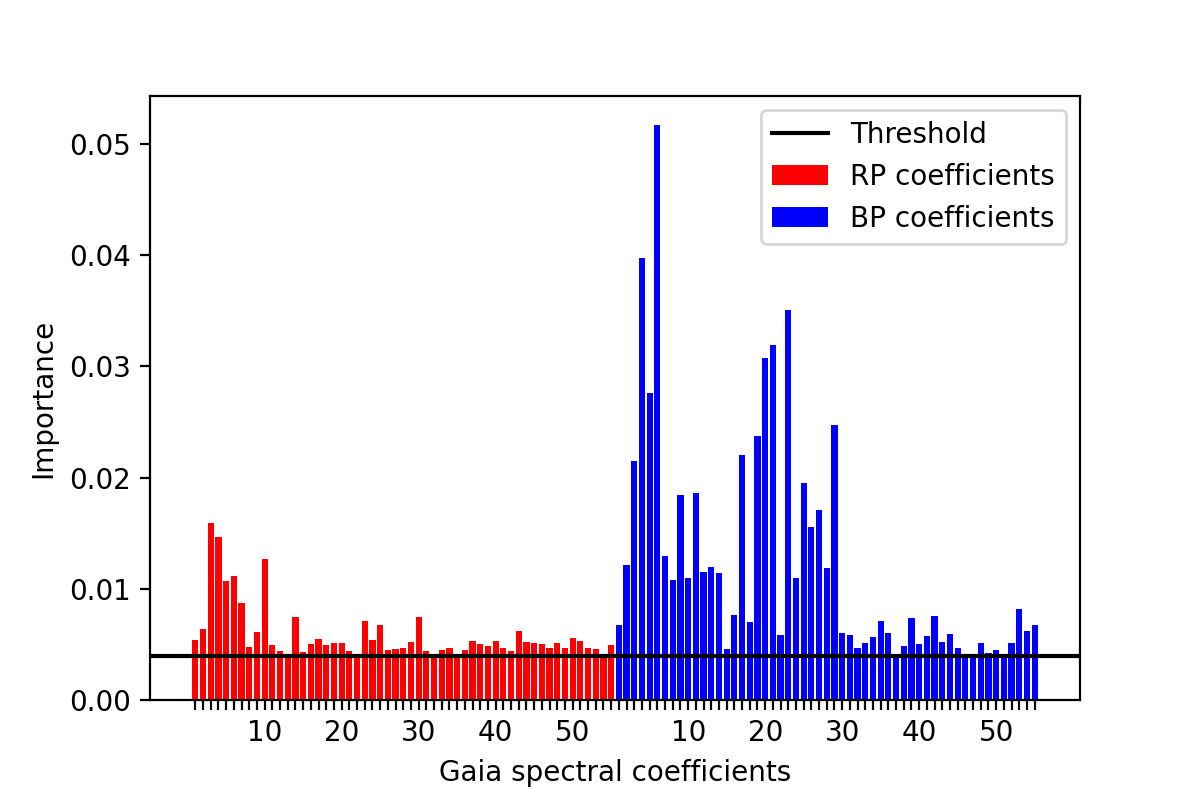}
    
    \caption{Feature importance as a function of the {\it Gaia} spectral coefficients for DA vs non-DA classification (top panel) and the non-DA classification into the different spectral subtypes (bottom panel). An importance threshold of 0.8\% is represented by a black horizontal line.}   
\label{f:Features}
\end{figure}

\section{Feature importance}
\label{s:fea}

As an ensemble learning method, the Random Forest algorithm constructs multiple decision trees combining their predictions to achieve the more accurate and stable result. In this construction, some features (variables or parameters of the sample) play a more remarkable role than others. Even more, one can remove some features without significantly altering the result. In our case, the features are the 110 {\it Gaia} spectral coefficients. We aim to analyse which of them have the highest importance for each classification. The method used to compute the feature importance was the mean decrease in impurity (MDI), which is based on the decrease of node impurity averaged over the whole Random Forest. This can be understood as follows. When a decision tree is generated, decision nodes are created. Node impurity is a measurement of the amount of classes in a certain decision node. They are said to be pure if they only comprise one class. Therefore, the most important features in our analysis are the ones that reduce the node impurity the most across the forest. These will, of course, be dependent on the set that is being classified. For instance, the coefficients that rule the Balmer lines are capital in a DA non-DA classification, but of no importance in a non-DA classification.

In Figure \ref{f:Features} we show the feature importance obtained by the MDI method as a function of the {\it Gaia} spectral coefficients for the DA versus non-DA (left panel) and the non-DA subtype (right panel) classification. 

\begin{figure*}[ht]
        \includegraphics[width=0.48\textwidth,trim=-100 0 0 0, clip]{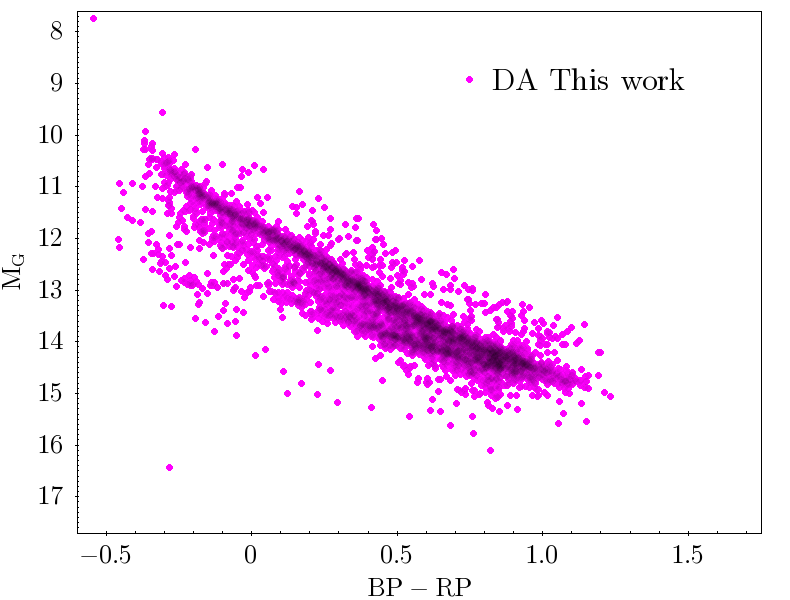}
            \includegraphics[width=0.48\textwidth,trim=0 0 -100 0, clip]{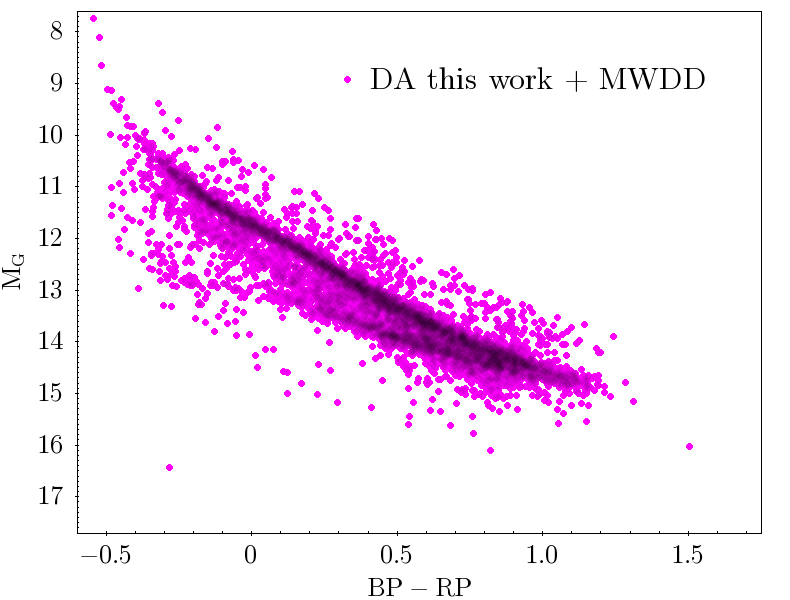}   
            
        \includegraphics[width=0.48\textwidth,trim=-100 0 0 0, clip]{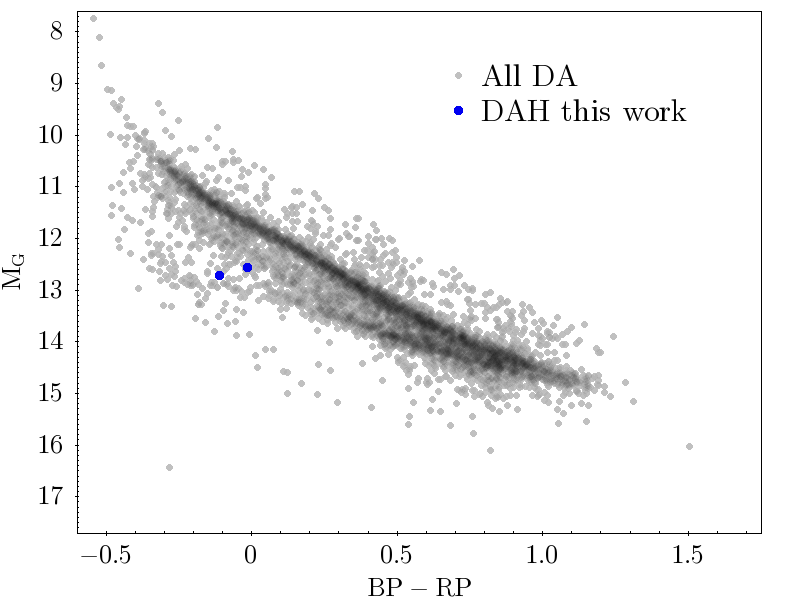}
            \includegraphics[width=0.48\textwidth,trim=0 0 -100 0, clip]{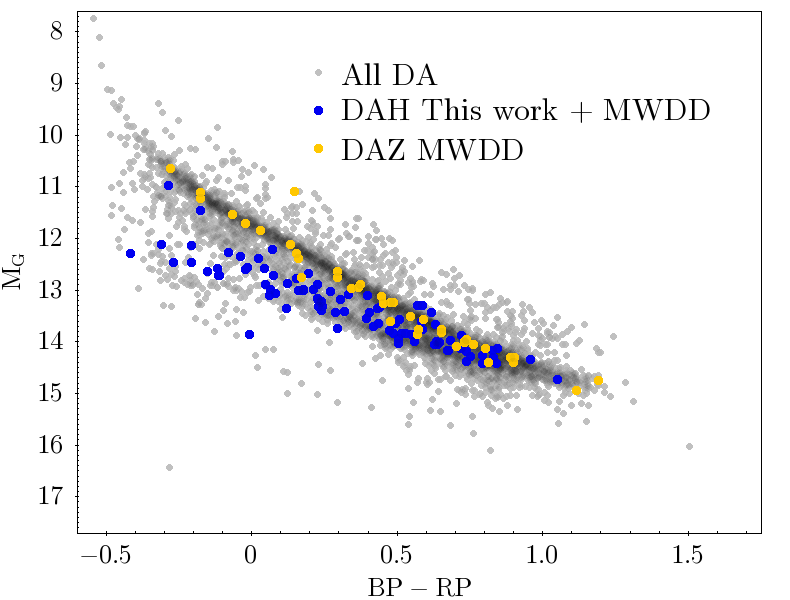}
 \caption{{\it Gaia} HR diagrams showing DA white dwarfs. {\sl Left panels:} DA white dwarfs classified in this work. {\sl Right panels:} entire population of DA white dwarfs (i.e. those classified in this work and those labelled in MWDD).}
    \label{f:HR_DAs}
\end{figure*}

Regarding the DA versus non-DA classification, the most important coefficients are approximately the 15 first red coefficients and the 20 first blue coefficients.  Moreover, if we consider a 0.8\% threshold (marked as black line in Fig \ref{f:Features}), we can eliminate most of the low-significant spectral coefficients, representing the remaining ones 73.6\% of the information.  

This result implies that a greater importance is placed in the BP information; indeed, all Balmer lines except {H}${\alpha}$, \ion{He}{I} lines, Swan bands and most metallic lines fall in the BP wavelength range. The most important feature, however, corresponds to the RP. We identify it with the Balmer {H}${\alpha}$ line. Since DAs show H features, it is predictable that the algorithm considers this spectral line as the most important to distinguish between DAs and non-DAs. 

With respect to the non-DA classification into its spectral subtypes, the feature importance distribution (right panel) reveals that blue coefficients are the most relevant. Applying the same 0.8\% threshold that in the previous case, approximately the first 30 coefficients contain the 52\% of the information. As most of the type-characteristic spectral lines (for instance, most \ion{He}{I} lines, the Swan bands, or \ion{Ca}{II} lines) appear in the wavelength range covered by the BP, rather than in the RP range, this result is both expected and consistent with our previous knowledge.

\section{The {\it Gaia} 100-pc sample classification summary}
\label{s:100pc}

In this section, we present a summary of our white dwarf spectral classification. From the 9\,446 classified white dwarfs, 4\,737 have been classified as DA, 2 as DAH, 76 as DB, 4\,437 as DC, 62 as DQ and 132 as DZ. The original, labelled MWDD sample used as training comprises 2\,905 objects; 1\,845 DA, 90 DAH, 97 DB, 573 DC, 117 DQ and 125 DZ. 

Consequently, the number of classified objects within 100 pc from the Sun has been increased by 257\% for DAs, 2.2\% for DAHs, 78.4\% for DBs, 774\% for DCs, 53\% for DQs and 105.6\% for DZs.

\begin{figure*}[pht]
        \includegraphics[width=0.48\textwidth,trim=-100 0 0 0, clip]{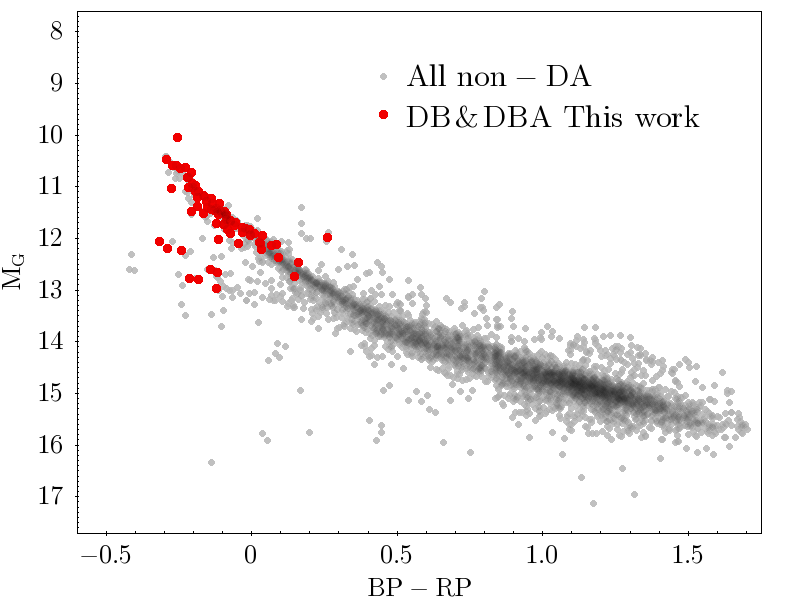}
            \includegraphics[width=0.48\textwidth,trim=0 0 -100 0, clip]{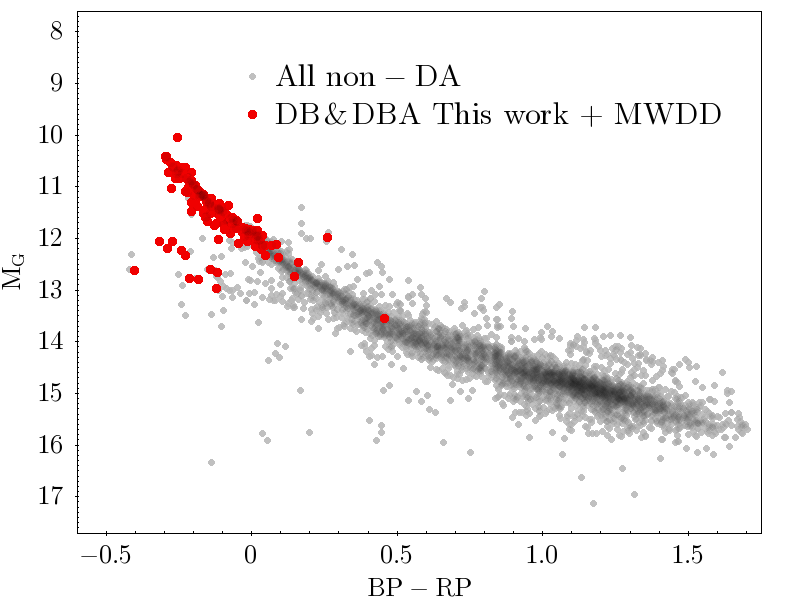}
            
        \includegraphics[width=0.48\textwidth,trim=-100 0 0 0, clip]{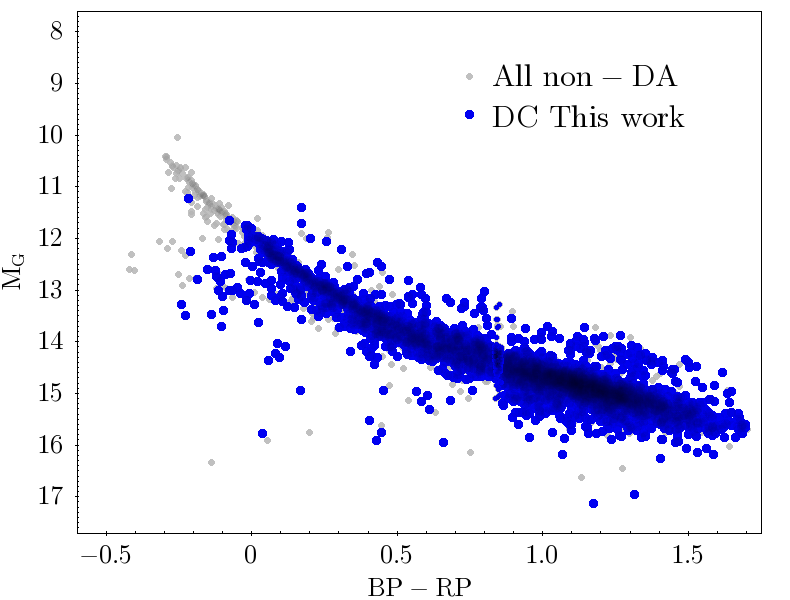}
            \includegraphics[width=0.48\textwidth,trim=0 0 -100 0, clip]{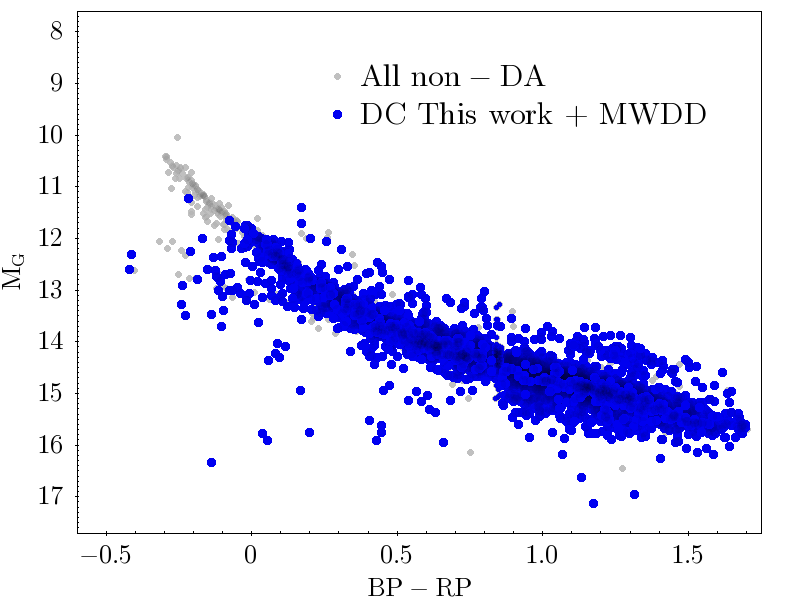}

        \includegraphics[width=0.48\textwidth,trim=-100 0 0 0, clip]{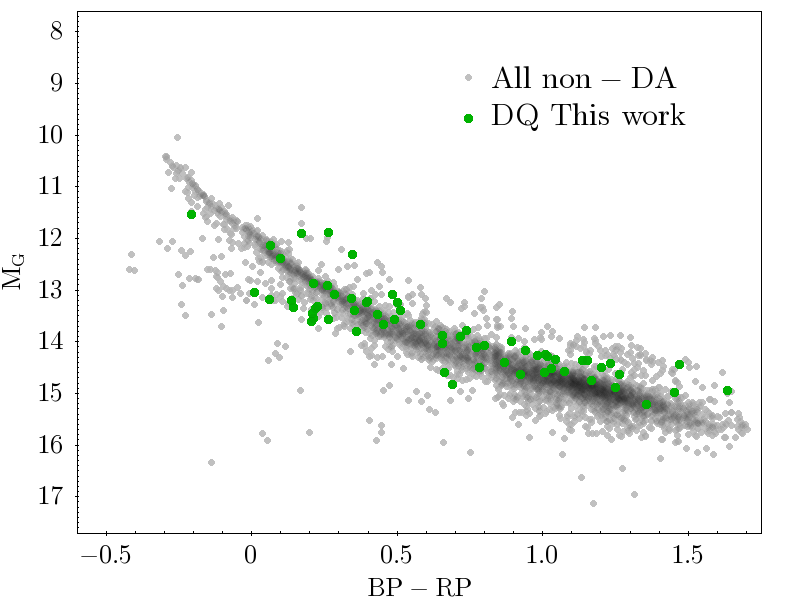}
            \includegraphics[width=0.48\textwidth,trim=0 0 -100 0, clip]{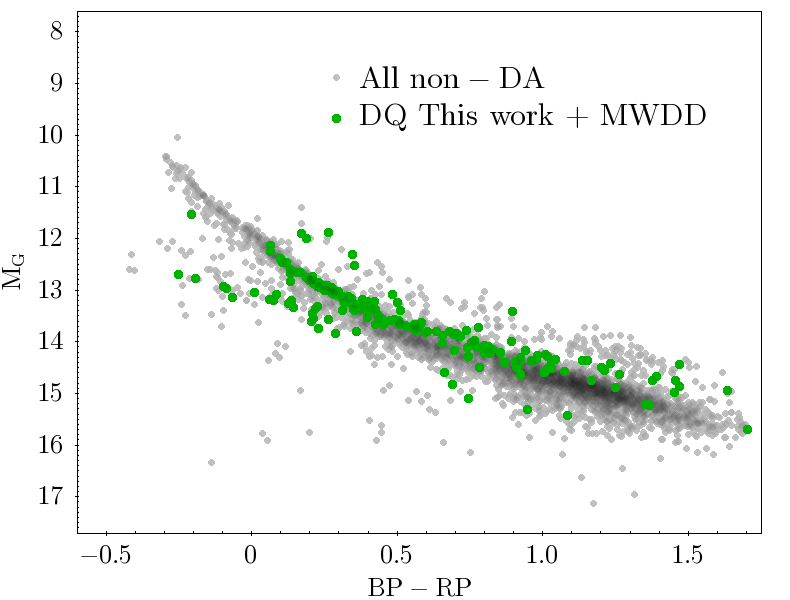}

       \includegraphics[width=0.48\textwidth,trim=-100 0 0 0, clip]{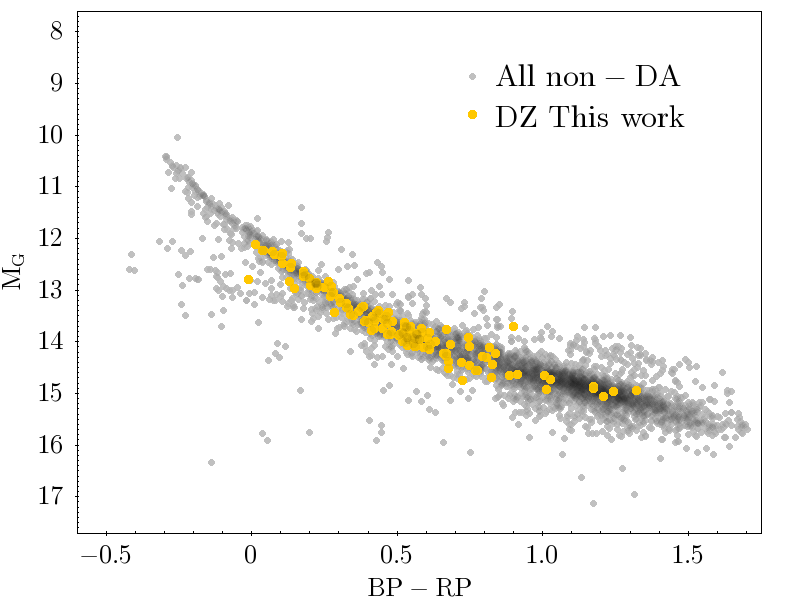}
            \includegraphics[width=0.48\textwidth,trim=0 0 -100 0, clip]{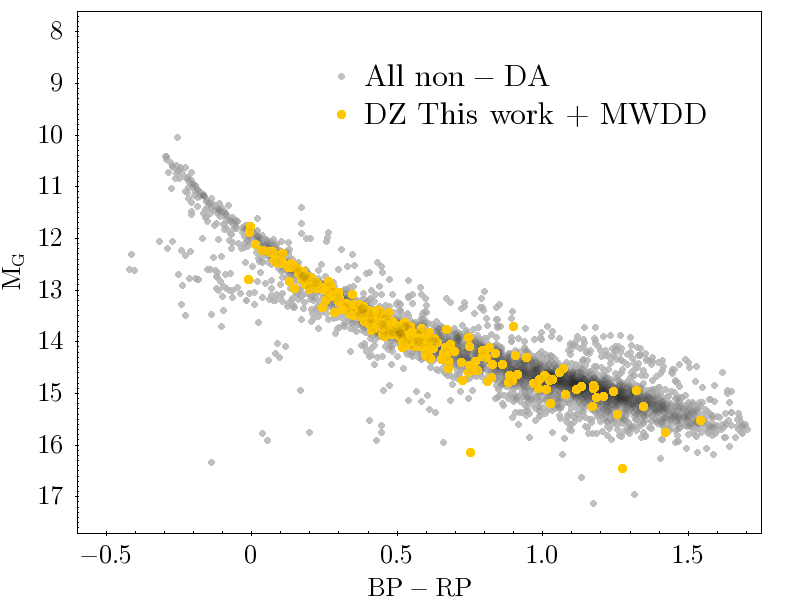}
                \caption{As Fig. \ref{f:HR_DAs} but for the different spectral subtypes of non-DAs.}
    \label{f:HR_nonDAs}
\end{figure*}

Figure \ref{f:HR_DAs}  show the \G HR diagrams for the white dwarfs classified in this work as DA and its secondary types (left panels), while in Figure  \ref{f:HR_nonDAs} are represented the corresponding  for objects classified as DB, DC, DQ and DZ by our algorithm (left panels). For completeness, we also show the HR diagrams  including those white dwarfs previously classified in the MWDD (right panels).

Additionally, all the objects studied here are collected in a list, where we provide their corresponding spectral classification among other {\it Gaia} parameters. A representative excerpt of this catalogue is presented in Table \ref{tab:catalogue}. The whole catalogue can be found in the electronic version of the paper. Moreover, for illustrative purposes, in Annex \ref{s:ann} we show some examples of  {\it Gaia} spectra corresponding to white dwarfs of different spectral types classified by our algorithm. These spectra are compared to the {\it Gaia} spectra of white dwarfs labelled in the MWDD.

\begin{table*}[p]
    \caption[]{{\it Gaia} 100 pc white dwarf sample catalogue classified by our Random Forest algorithm into spectral types.}
    \label{tab:catalogue}
\begin{center}
    \begin{tabular}{lcccccc}
            \noalign{\smallskip}
            \hline
            \noalign{\smallskip}
        \G Source ID & RA & Dec & $M_G$ & BP-RP & $d$ & $S_{\rm type}$ \\
            & (deg) & (deg) & (mag) & (mag) & (pc) & \\
            \noalign{\smallskip}
            \hline
            \noalign{\smallskip}
        6617996741403360128 & 328.75 & -27.84 & 14.42 & 0.91 & 23.5 & DA\\
        3022956969731332096 & 88.79 & -4.18 & 15.19 & 1.13 & 6.44 & DA \\
        5314177402013456256 & 141.20 & -49.26 & 14.59 & 0.89 & 22.57 & DA \\
        863131372427958912 & 176.43 & 63.10 & 14.29 & 0.92 & 24.01 & DA \\
        4789516154317811456 & 65.99 & -45.85 & 14.29 & 0.89 & 29.94 & DA \\
        3501922067493606272 & 109.10 & -23.30 & 13.24 & 0.88 & 41.89 & DA \\
        5142336825646176256 & 25.67 & -17.24 & 14.36 & 0.90 & 40.05 & DA \\
        6906197808698867712 & 306.88 & -7.60 & 14.63 & 1.00 & 51.38 & DA \\
        5058635403471767680 & 49.30 & -29.19 & 14.74 & 0.96 & 31.50 & DA \\
        5432789383518999168 & 144.40 & -38.87 & 14.27 & 0.86 & 34.50 & DA \\
                  $\ldots$  & $\ldots$ & $\ldots$ & $\ldots$ & $\ldots$ & $\ldots$ & $\ldots$ \\
        5945252202546434432 & 266.95 & -52.12 & 12.72 & -0.11 & 38.92 & DAH \\
        3630648387747801088 & 204.92 & -7.22 & 12.57 & -0.01 & 56.33 & DAH \\
                  $\ldots$  & $\ldots$ & $\ldots$ & $\ldots$ & $\ldots$ & $\ldots$ & $\ldots$ \\
                          6600365522695593472 & 336.84 & -34.19 & 10.81 & -0.22 & 53.20 & DB \\
        6570892323240774144 & 329.51 & -43.47 & 11.10 & -0.19 & 50.59 & DB \\
        2446993162322393088 & 354.82 & -4.41 & 11.41 & -0.15 & 89.28 & DB \\
           $\ldots$  & $\ldots$ & $\ldots$ & $\ldots$ & $\ldots$ & $\ldots$ & $\ldots$ \\
        2628943473222829440 & 339.03 & -1.68 & 12.74 & 0.15 & 39.02 & DBA \\
        3375135698070213632 & 93.44 & 20.84 & 11.98 & 0.26 & 57.98 & DBA \\
        5657351404992422784 & 148.25 & -26.96 & 10.97 & -0.19 & 83.64 & DBA \\
        5719723160586005888 & 124.65 & -18.55 & 11.53 & -0.17 & 87.66 & DBA \\
        1608497864040134016 & 216.90 & 53.81 & 11.43 & -0.14 & 51.54 & DBA \\
        184735992329821312 & 75.43 & 33.40 & 11.72 & -0.10 & 74.98 & DBA \\
        5099182265566440320 & 52.15 & -8.26 & 11.49 & -0.21 & 87.54 & DBA \\
                  $\ldots$  & $\ldots$ & $\ldots$ & $\ldots$ & $\ldots$ & $\ldots$ & $\ldots$ \\
        4647914239368623616 & 87.62 & -76.81 & 14.63 & 1.09 & 85.91 & DC \\
        4983839647522981504 & 21.02 & -42.68 & 13.36 & 0.72 & 18.45 & DC \\
        4108828945319007744 & 256.67 & -26.73 & 14.73 & 0.72 & 13.05 & DC \\
        1196295211098415616 & 237.68 & 16.06 & 14.41 & 0.91 & 91.01 & DC\\
        4923825240566074624 & 3.01 & -54.02 & 15.37 & 1.36 & 73.78 & DC\\
                  $\ldots$  & $\ldots$ & $\ldots$ & $\ldots$ & $\ldots$ & $\ldots$ & $\ldots$  \\
        5717278911884258176 & 115.09 & -17.42 & 13.16 & 0.34 & 9.15 & DQ \\
        5332606522595645952 & 176.46 & -64.84 & 13.09 & 0.29 & 4.64 & DQ  \\
        946030529073021440 & 107.56 & 37.67 & 13.58 & 0.49 & 24.37 & DQ \\
        6797171060323993728 & 305.11 & -30.45 & 12.39 & 0.10 & 17.46 & DQ \\
        1932612039116771456 & 341.64 & 40.41 & 11.53 & -0.21 & 48.79 & DQ\\
        51090628850761088 & 57.83 & 19.63 & 14.29 & 1.02 & 79.04 & DQ \\
        6812238900812142720 & 329.21 & -24.88 & 14.41 & 0.87 & 67.19 & DQ \\
        2201461976641376512 & 335.06 & 60.36 & 14.25 & 1.01 & 94.86 & DQ\\
        3030820432081929088 & 115.74 & -12.80 & 14.45 & 1.47 & 76.83 & DQ\\
                          $\ldots$  & $\ldots$ & $\ldots$ & $\ldots$ & $\ldots$ & $\ldots$ & $\ldots$ \\
                5271072526109138176 & 124.31 & 124.31 & 15.21 & 1.36 & 38.88 & DQpec \\
                  $\ldots$  & $\ldots$ & $\ldots$ & $\ldots$ & $\ldots$ & $\ldots$ & $\ldots$ \\
        6054148143441683072 & 184.36 & -63.50 & 13.17 & 0.30 & 37.52 & DZ\\
        3327488430402704000 & 96.91 & 10.04 & 12.88 & 0.22 & 52.91 & DZ\\
        6598883720324218240 & 335.91 & -34.64 & 12.98 & 0.15 & 57.02 & DZ \\
        2645295955612242688 & 350.31 & 1.04 & 15.07 & 1.21 & 61.06 & DZ \\
        6429838617917056512 & 306.32 & -63.40 & 14.25 & 1.01 & 61.70 & DZ \\
        2067245446933399936 & 305.98 & 39.44 & 14.92 & 1.01 & 67.25 & DZ \\
        6505113009316709760 & 339.00 & -55.82 & 14.73 & 1.03 & 31.91 & DZ\\
        4823007896975577088 & 81.84 & -34.46 & 14.91 & 1.17 & 65.17 & DZ\\
        4217729529025009024 & 306.73 & -6.43 & 12.56 & 0.13 & 51.72 & DZ \\
                  $\ldots$  & $\ldots$ & $\ldots$ & $\ldots$ & $\ldots$ & $\ldots$ & $\ldots$ \\
            \noalign{\smallskip}
            \hline
    \end{tabular}
\end{center}
\end{table*}

\section{Comparison to other automatic classification methods.}
\label{s:others}

To assess the quality of the performance of our Random Forest algorithm, in this section, we compare it with other similar automated classification methods described in the literature. In particular, we analyze the results obtained by \cite{Olivier2023} in their white dwarf spectral classification using neural networks. 

\begin{figure}[ht]
    \includegraphics[width=1.0\columnwidth,trim=20 0 30 20, clip]{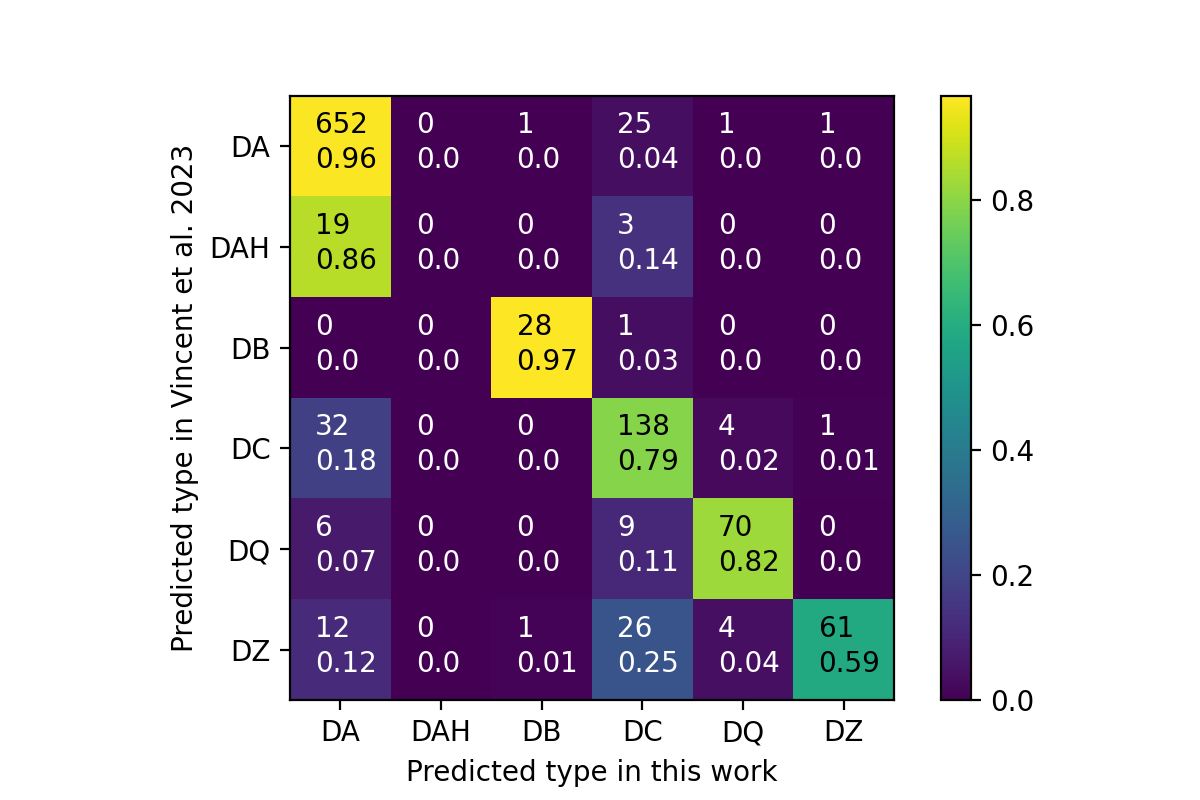}
                \caption{Confusion matrix of objects that appear both in our classification and in \cite{Olivier2023}.}
    \label{f:cmatrix_cross}
\end{figure}

Although the methodology is not the same, and neither the classification sample, the training sample, nor the input data are identical, we can establish a certain comparative analysis of the results. For instance, our work is focused on the {\it Gaia} 100 pc white dwarf sample for mainly  primary spectral types (DA, DB, DC, DQ and DZ) classification, while the work by  \citet{Olivier2023} consists in a more general approach for white dwarf candidate selection and spectroscopic classification. This includes primary spectral types, and also other subtypes, such as DO, hot DQ, DAH, PG 1159 objects and various types of subdwarfs, as well as white dwarfs plus main sequence binaries. Moreover, the input data used in \cite{Olivier2023} comes mainly from both the {\it Gaia} parameter database and Sloan Digital Sky Survey spectra, while in our study, we only focused on the Hermite coefficients from {\it Gaia} spectra.

Nevertheless, for comparative purposes, we have constructed a confusion matrix with the spectral classification of the objects that appear in both the present work and \cite{Olivier2023}. Both catalogues were cross-matched and $1\,103$ objects were found in both tables. Of them, six were classified in \cite{Olivier2023} as cataclismic variables and two as hot subdwarf stars; these objects were disregarded in the construction of the confusion matrix. As such, the resulting confusion matrix, shown in Figure \ref{f:cmatrix_cross}, contains 1\,095 objects.

The obtained confusion matrix is nearly diagonal, which indicates a general good agreement (86\% accuracy considering the \citet{Olivier2023} classification as true labels). The only remarkable exception are magnetic DAH white dwarfs. For the 22 white dwarfs classified as DAH in \cite{Olivier2023}, 19 of them are classified by our algorithm as DAs and 3 as DCs. In Section \ref{3.3.1} we showed that our algorithm is practically unable to distinguish DAs from DAHs. The reason is due to the lack of the necessary {\it Gaia} spectra resolution to resolve the fine magnetic splitting. Therefore, this result for DAHs is entirely understandable.

Similarly, but to a lesser extent, the 104 DZ candidates in \cite{Olivier2023} are broadly in agreement with our classification (61 objects, 59\%), but our algorithm classifies the discrepancies as DC, DA, or DQs. Once again, the low-resolution of {\it Gaia} spectra prevent to more accurate classification. 

We can conclude that, despite the quality of the input data (which, obviously, the better the quality data, the better the performance), our Random Forest algorithm is a feasible tool with very low-cost computer time consuming and model-independent tuning parameters, allowing a reliable and robust classification of white dwarf spectra

\section{Conclusions}
\label{s:conc}

By using Artificial Intelligence techniques based on a Random Forest algorithm, we have analyzed the information contained in the coefficients of {\it Gaia} spectra. Even though these spectra are of low resolution, we have verified their usefulness in classifying the population of white dwarfs into their different spectral types. In particular, we have classified the full 100 pc {\it Gaia} white dwarf population into their primary spectral types (i.e. DA, DB, DC, DZ and DQ) and also finding some secondary types (DAH, DBA and DQpec). 

A summary of the main findings is as follows:
   \begin{enumerate}
      \item The Random Forest algorithm is able to classify DA and DC white dwarfs with excellent recall (>97\%), DBs with very good recall (>80\%), and DQs and DZs with improvable recall (<50\%).
      \item In spite of the low recall, DQ and DZ white dwarfs are classified with an excellent precision (>90\%).
      \item While the algorithm performance is certainly improvable at correct identification of DQ and DZ white dwarfs, its high precision for these spectral types, as well as DB, allows us to use the classifying algorithm as a white dwarf finder for these specific types.
      \item With the possible exception of the DBA and DQpec subtypes, spectral subtypes do not seem to be recognised by the algorithm. Low resolution inherent to {\it Gaia} mean spectra seems to be the limiting factor for classification, as  non-prominent spectral lines are not expected to be detected in them.
      \item Our algorithm has identified 76 DB (most of them, 60, DBA), 60 DQ (9 of them DQpec), 132 DZ and 2 DAH candidates in a 100-pc radius around the Sun. For comparison, the MWDD classified sample used in validation tests and as training material contained 117 DQ and 125 DZ white dwarfs.
   \end{enumerate}

In conclusion, this initial classification of the entire white dwarf population within 100 pc opens the door to more precise studies of mass distribution and luminosity function, among others, based on the spectral classification of these objects. In parallel, we have initiated a spectroscopic follow-up of a large sample of candidate objects to confirm their classification.

\begin{acknowledgements}
We acknowledge support from MINECO under the PID2020-117252GB-I00 grant and by the AGAUR/Generalitat de Catalunya grant SGR-386/2021. EMGZ  also acknowledges financial support from Banco de Santander, under a Becas Santander Investigación/Ajuts de Formació de Professorat Universitari (2022\_FPU‐UPC\_16) grant.
\end{acknowledgements}


\bibliographystyle{aa}
\bibliography{RFnDA}

\section{Annex}
\label{s:ann}
In this Annex, we show examples of {\it Gaia} white dwarf spectra classified by our Random Forest algorithm. In all of them, the expected location of Balmer spectral lines are shown, as well as some characteristic spectral lines for every spectral type. These include \ion{He}{I} spectral lines for DB white dwarfs, Swan bands for DQs and a selection of metallic lines for DZs, detailed in subsection \ref{9.6}. For comparative purposes, we accompanied each classified spectrum obtained by our algorithm with one spectrum corresponding to a white dwarf classified with the same type from the  MWDD. 

All shown spectra have been obtained from {\it Gaia} internally calibrated spectra, using the GaiaXPy Python package  to transform them into wavelength-flux externally calibrated spectra. Oscillatory behaviour in the spectra, as explained in Section \ref{s:method} at the blue and red extremes, are characteristic of {\it Gaia} externally calibrated spectra, and are caused by the behaviour of the Hermite polynomials used to convey them near the extremes

\subsection{DA spectra}

\begin{figure*}[ht]
\centering
    \includegraphics[width=1.0\columnwidth,trim=-20 0 0 0, clip]{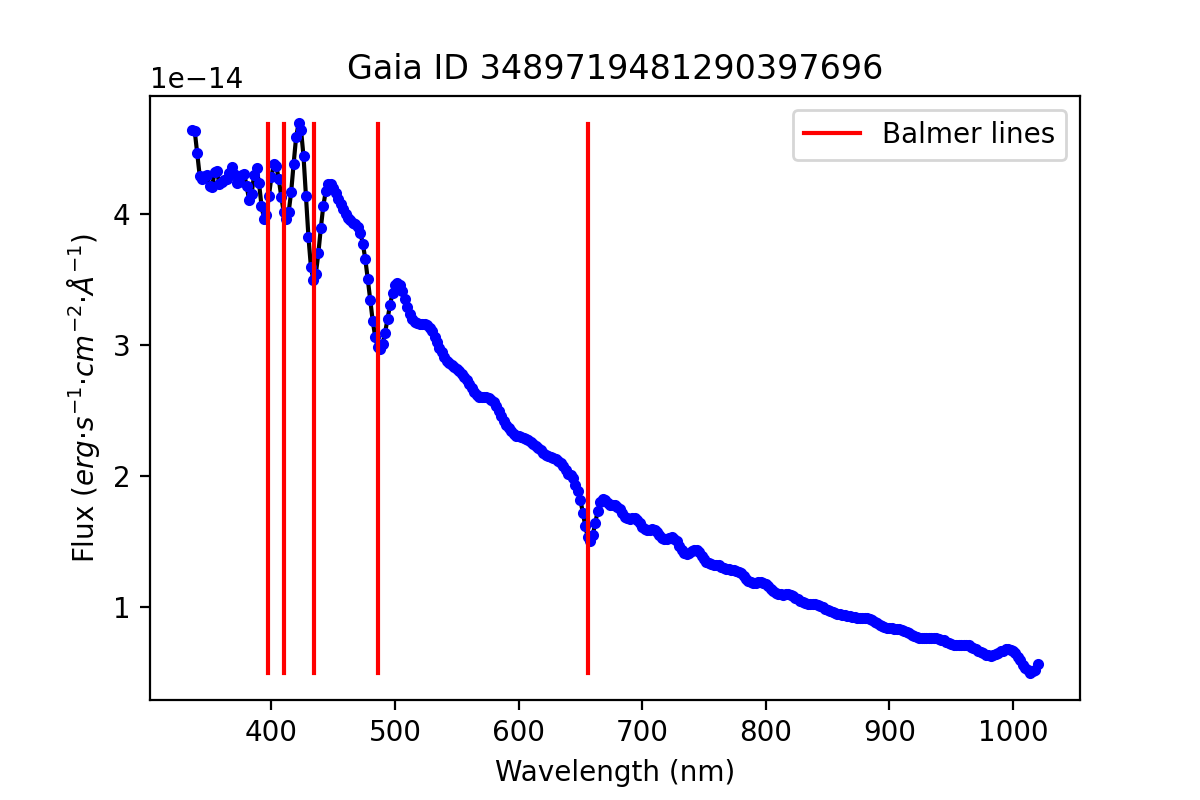}
    \includegraphics[width=1.0\columnwidth,trim=0 0 -20 0, clip]{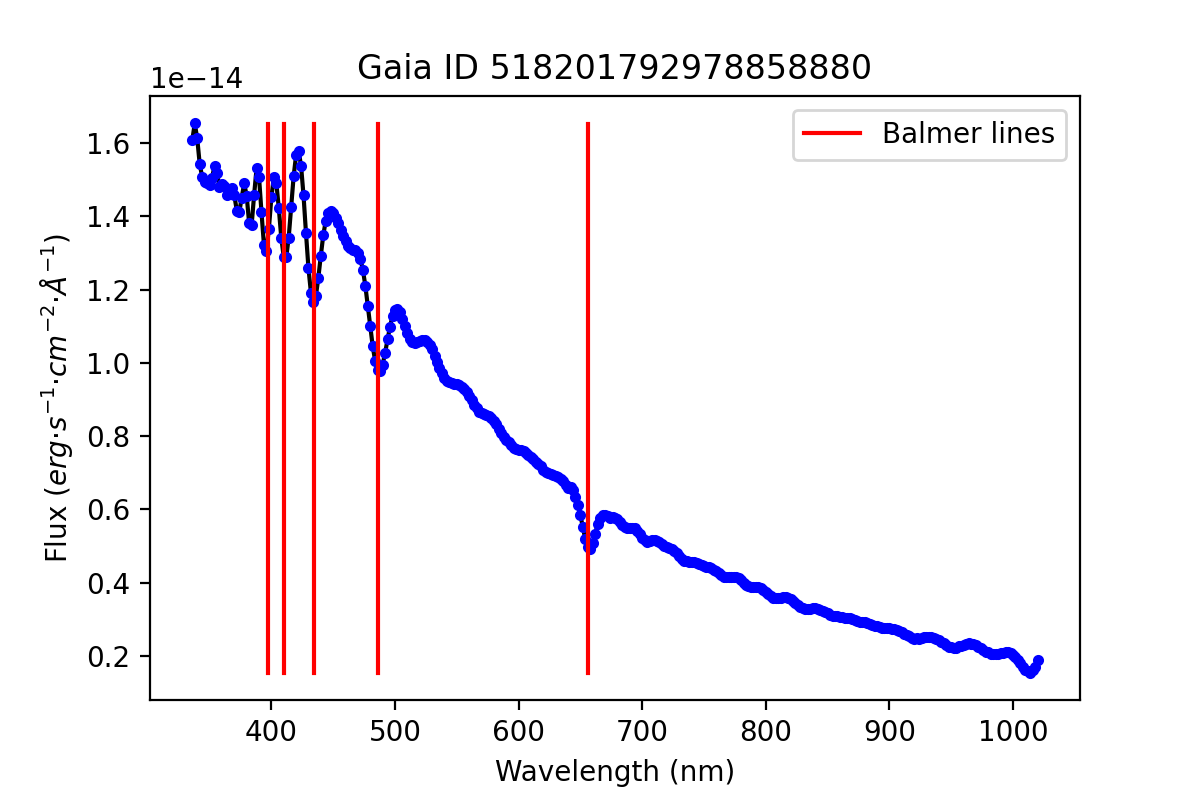}
    \caption{Examples of {\it Gaia} spectra. {\sl Left panel:} of a white dwarf classified as DA by MWDD. {\sl Right panel:} of a white dwarf classified as DA by our algorithm.}
    \label{f:DA_MWDD_RF}
        \vspace{0.5cm}
\end{figure*}

In this section, we show a DA spectrum from a registered white dwarf in the MWDD and a spectrum classified as such in this work in Figure \ref{f:DA_MWDD_RF}. Balmer lines (H$\alpha$, H$\beta$, H$\gamma$, H$\delta$ and H$\epsilon$) are marked in both spectra. These lines are very prominent, which explains the excellent recall for DAs (see confusion matrix in top panel of Figure \ref{f:cmvaltests}.)

\subsection{DAH spectra}

\begin{figure*}[ht]
\centering
    \includegraphics[width=1.0\columnwidth,trim=-20 0 0 0, clip]{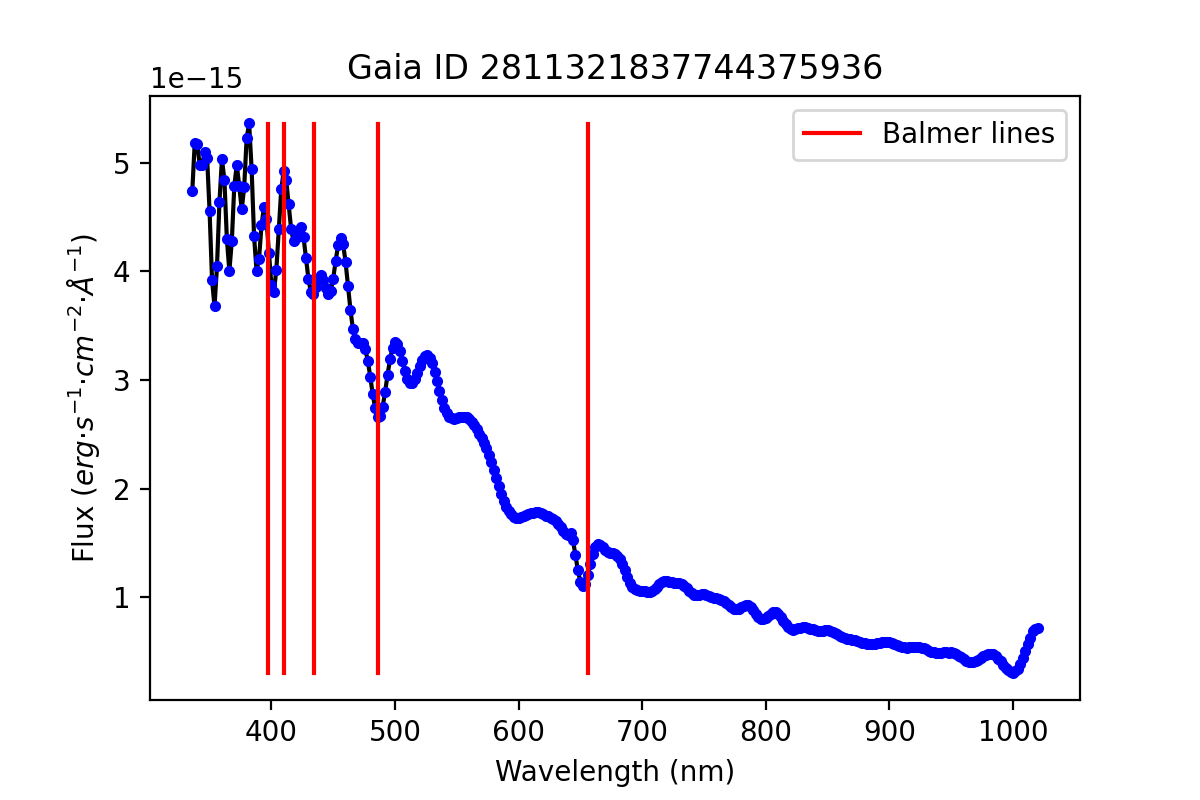}
    \includegraphics[width=1.0\columnwidth,trim=0 0 -20 0, clip]{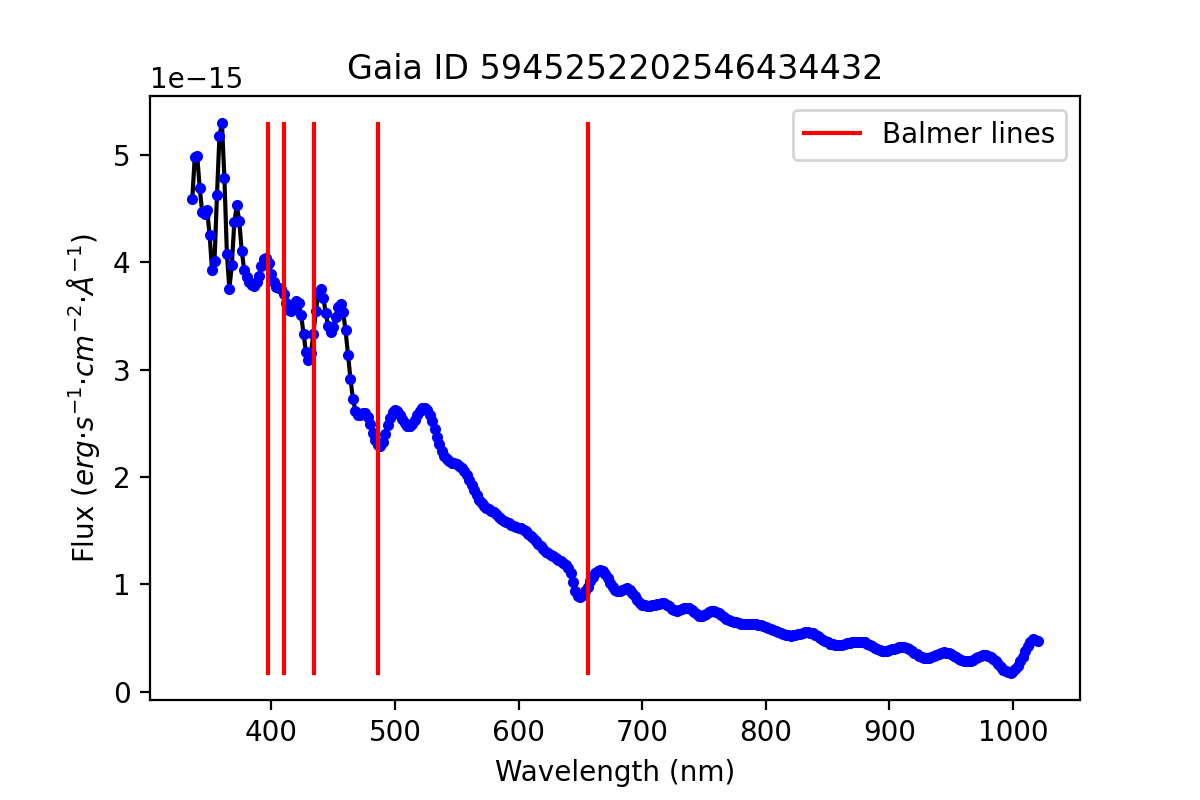}
    \caption{As Fig. \ref{f:DA_MWDD_RF} but for a DAH classified in MWDD (left panel) and  a DAH classified by our algorithm (right panel).}
    \label{f:DAH_MWDD_RF}
        \vspace{0.5cm}
\end{figure*}

{\it Gaia} spectra for those objects classified as DAH are shown in this section, in Figure \ref{f:DAH_MWDD_RF}. As hydrogen-dominated DAs, Balmer lines are prominent; Zeeman effect magnetic splitting is not noticeable in the objects labeled as such in MWDD nor in those identified by our Random Forest algorithm. This serves to illustrate the role of \G spectral low resolution in explaining the low recall for DAH white dwarfs  by our algorithm.

\subsection{DB spectra}

\begin{figure*}[ht]
\centering
    \includegraphics[width=1.0\columnwidth,trim=-20 0 0 0, clip]{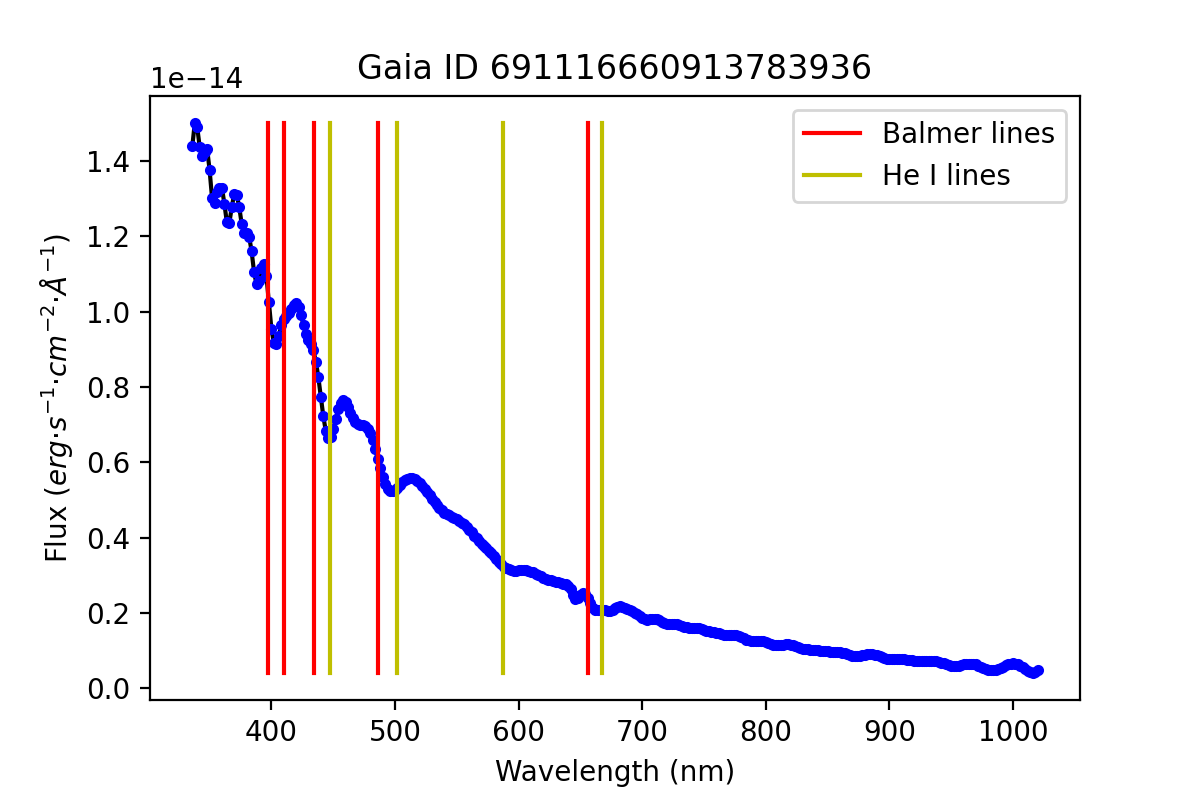}
    \includegraphics[width=1.0\columnwidth,trim=0 0 -20 0, clip]{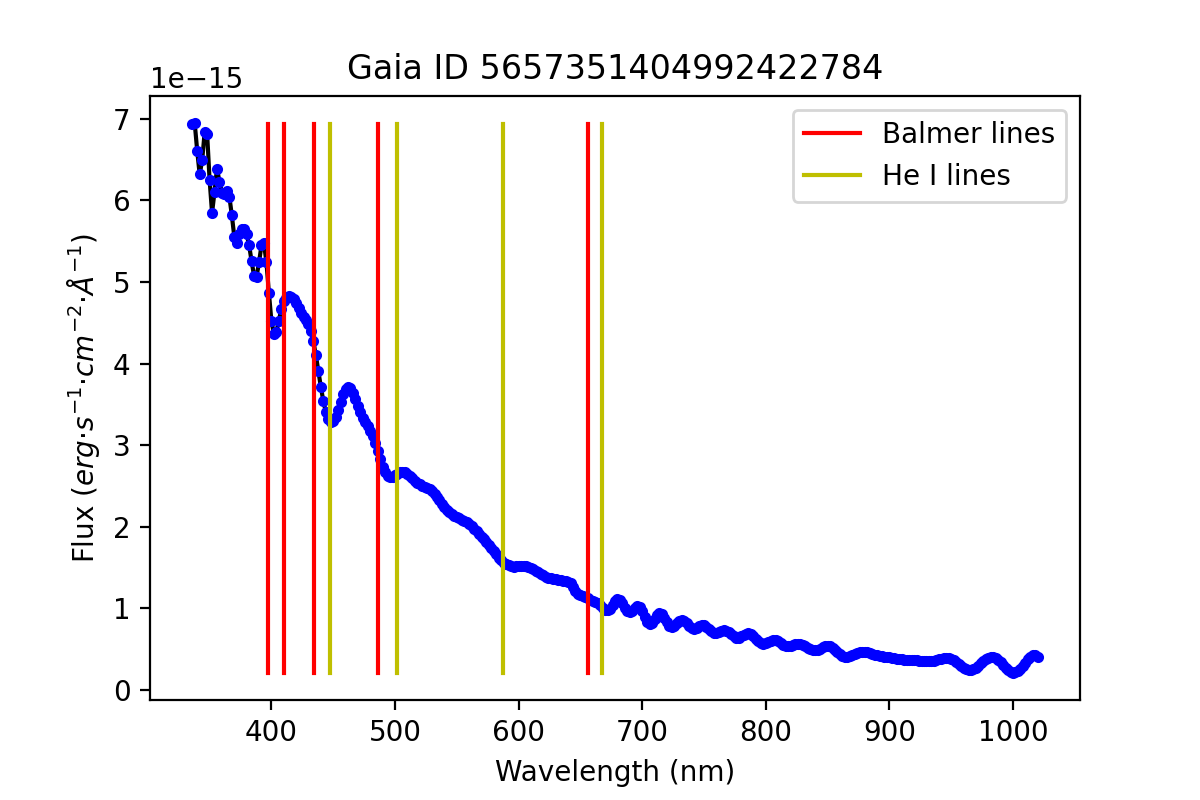}
    \caption{As Fig. \ref{f:DA_MWDD_RF} but for a DB classified in MWDD (left panel) and  a DB classified by our algorithm (right panel).}
    \label{f:DB_MWDD_RF}
    \vspace{0.5cm}
\end{figure*}

 The {\it Gaia} spectra of a white dwarf classified as DB by the MWDD and by our algorithm are presented in Figure \ref{f:DB_MWDD_RF}. Neutral helium lines are present and noticeable in the shown spectra, and the \ion{He} {I} lines at 4\,471, 5\,015, 5\,875 and 6\,678 $\AA$ have been marked.

\subsection{DC spectra}

\begin{figure*}[ht]
\centering
    \includegraphics[width=1.0\columnwidth,trim=-20 0 0 0, clip]{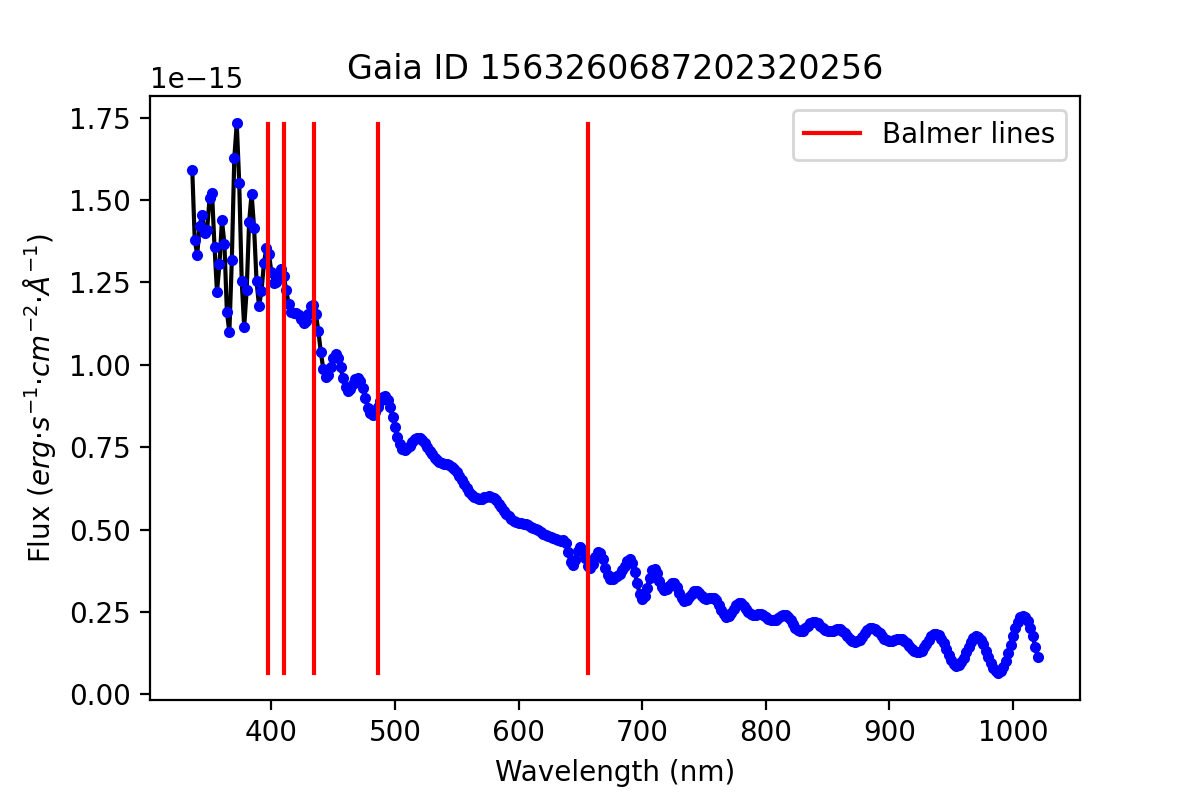}
    \includegraphics[width=1.0\columnwidth,trim=0 0 -20 0, clip]{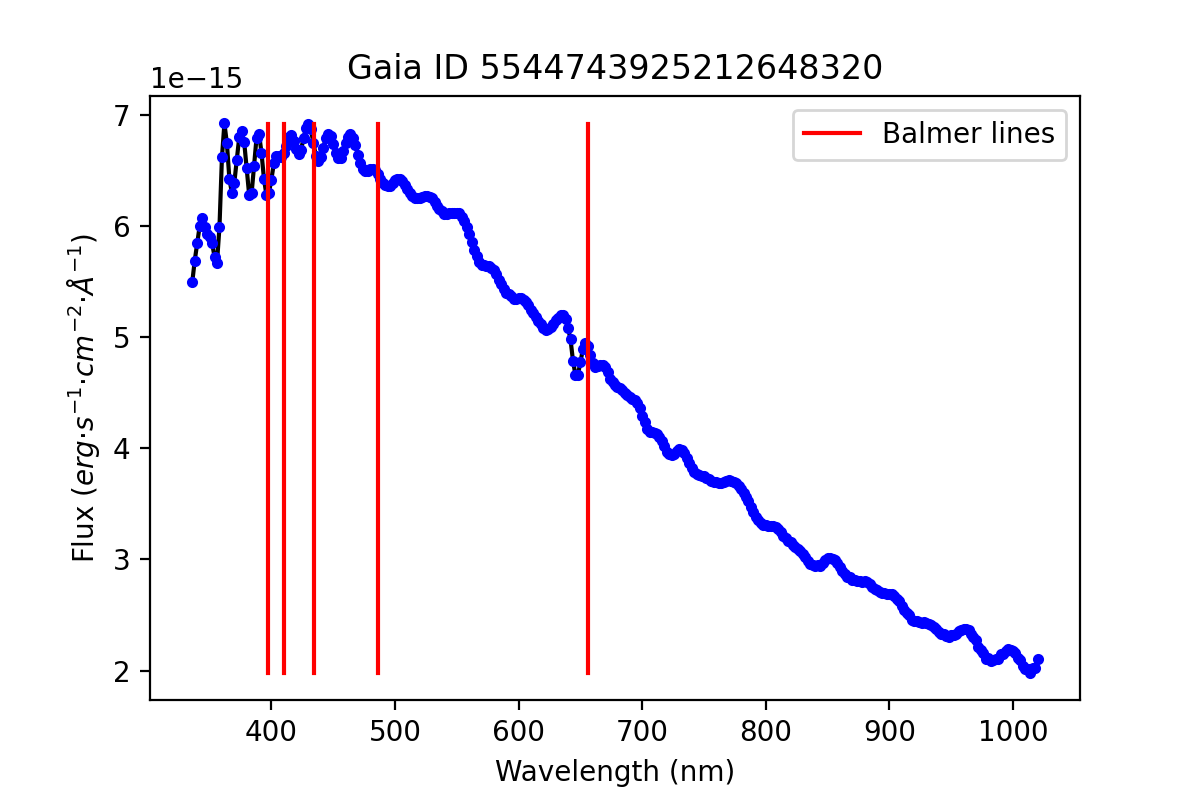}
    \caption{As Fig. \ref{f:DA_MWDD_RF} but for a DC classified in MWDD (left panel) and  a DC classified by our algorithm (right panel).}
    \label{f:DC_MWDD_RF}
        \vspace{0.5cm}
\end{figure*}

In Figure \ref{f:DC_MWDD_RF} we show the {\it Gaia} DC spectra of a MWDD and here classified white dwarf. It can clearly be visualized a characteristic, featureless spectra (except for oscillations caused by Hermite polynomials behaviour). No spectral lines could be found; wavelengths corresponding to \ion{H}{I} Balmer lines have been marked nonetheless to stress the featurelessness of these spectra.

\subsection{DQ spectra}

\begin{figure*}[ht]
\centering
    \includegraphics[width=1.0\columnwidth,trim=-20 0 0 0, clip]{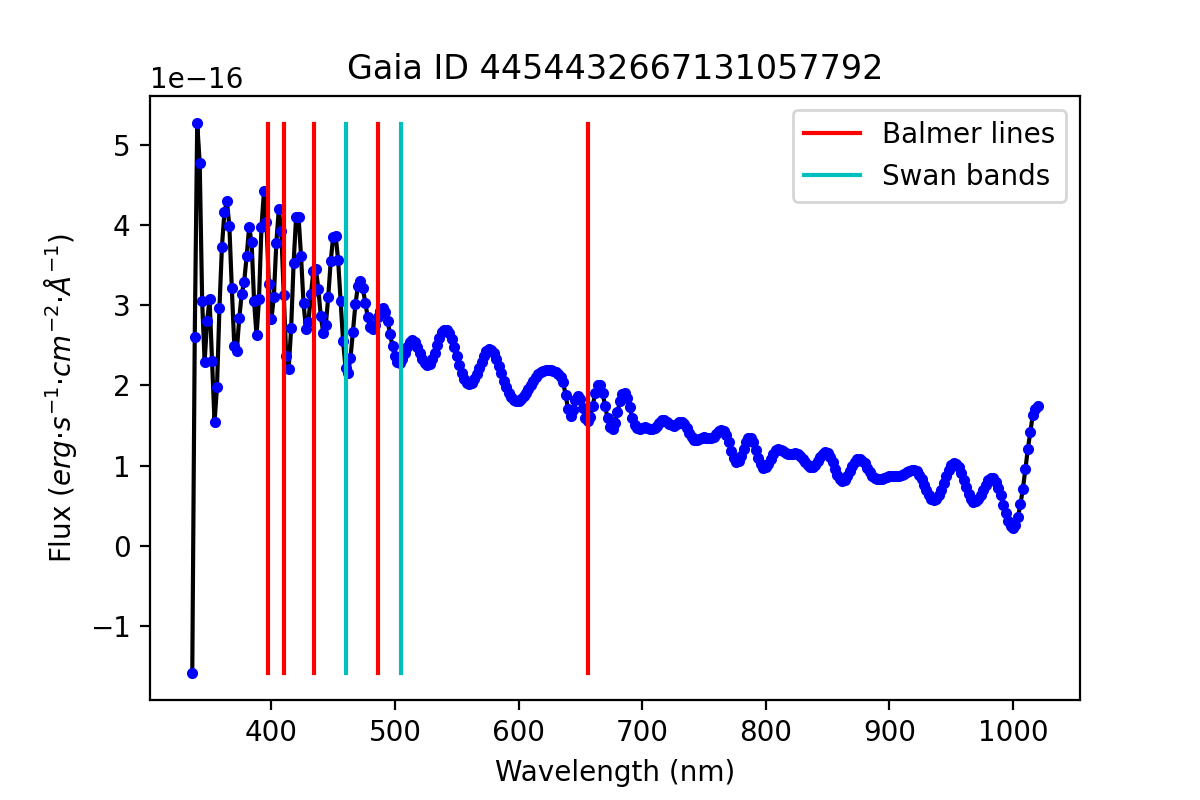}
    \includegraphics[width=1.0\columnwidth,trim=0 0 -20 0, clip]{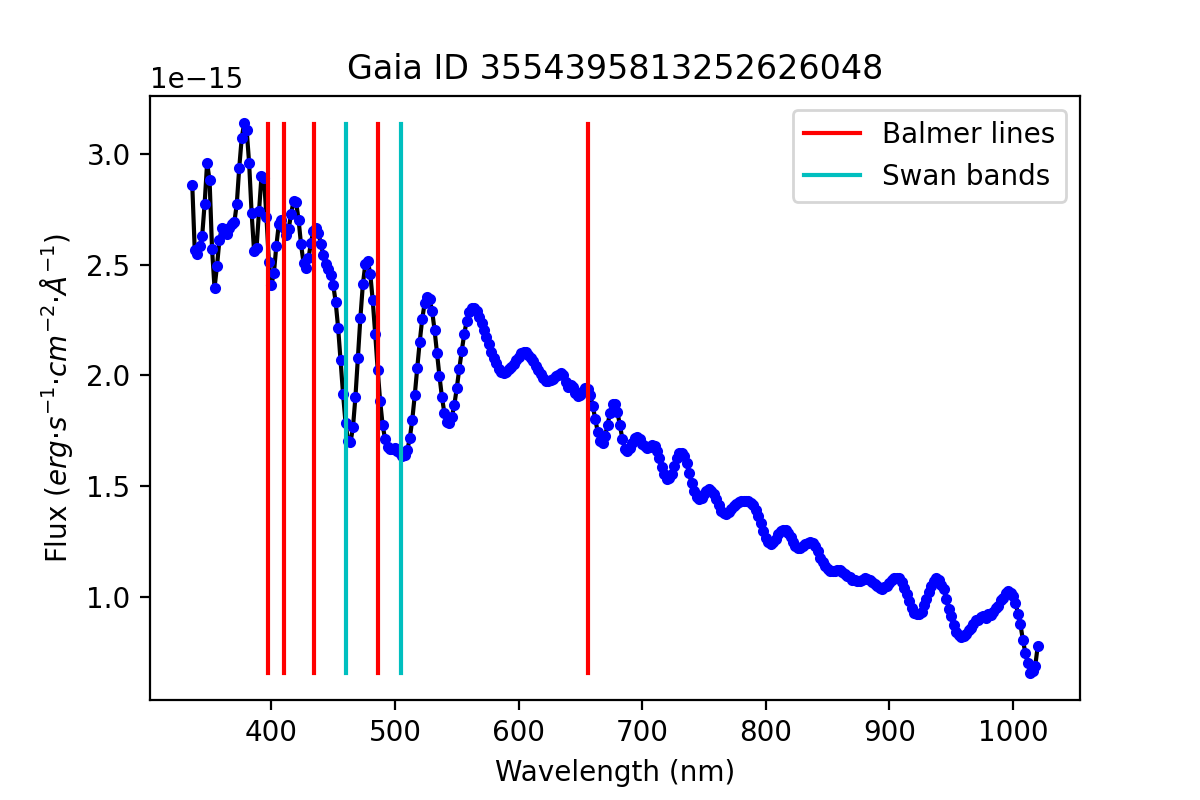}
    \caption{As Fig. \ref{f:DA_MWDD_RF} but for a DQ classified in MWDD (left panel) and  a DQ classified by our algorithm (right panel).}
    \label{f:DQ_MWDD_RF}
        \vspace{0.5cm}
\end{figure*}

The defining characteristic of DQ white dwarfs is the presence of carbon spectral lines. Most atomic carbon lines are outside of the {\it Gaia} BP and RP spectral coverage. However, Swan bands, vibrational bands characteristic of diatomic carbon (C$_{2}$) are in the visible spectra. These are marked in the shown spectra (Figure \ref{f:DQ_MWDD_RF}). Even though Swan bands comprise a high number of vibrational transitions, for clarity they have been marked at 4\,600 and 5\,050 $\AA$.

\begin{figure*}[ht]
\centering
    \includegraphics[width=1.0\columnwidth,trim=-20 0 0 0, clip]{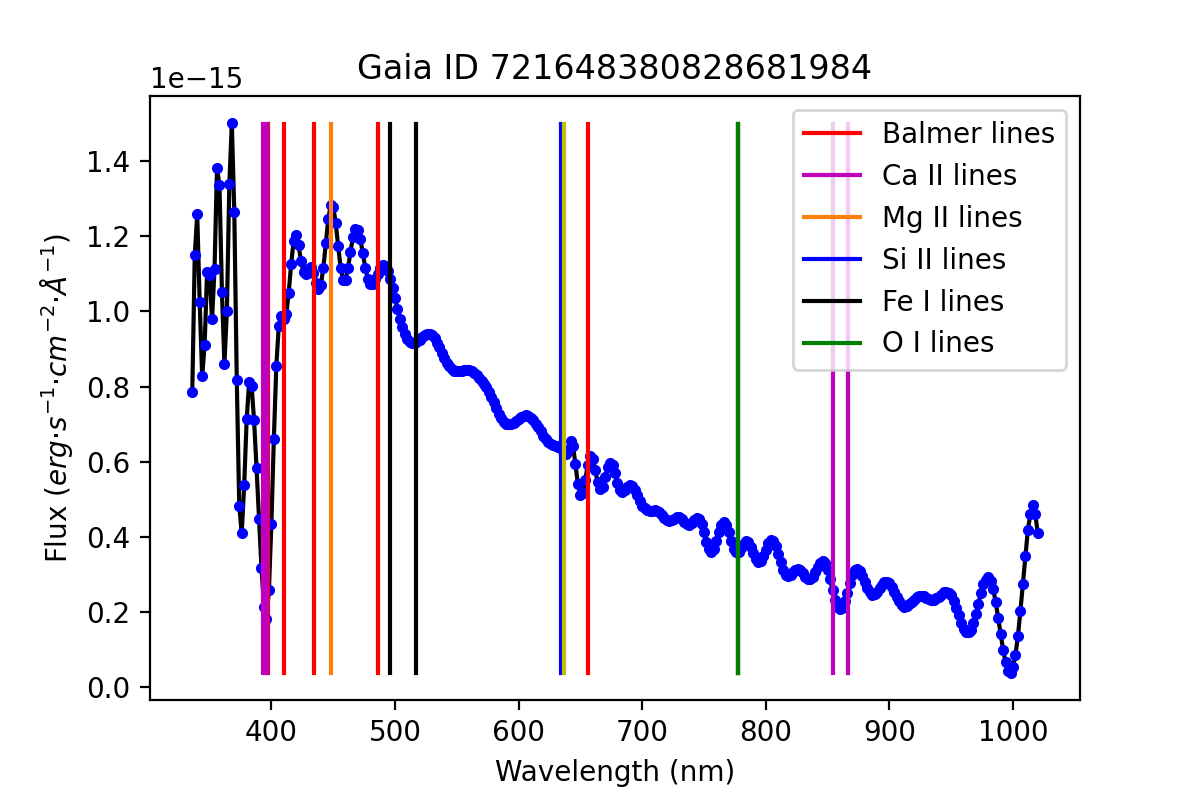}
    \includegraphics[width=1.0\columnwidth,trim=0 0 -20 0, clip]{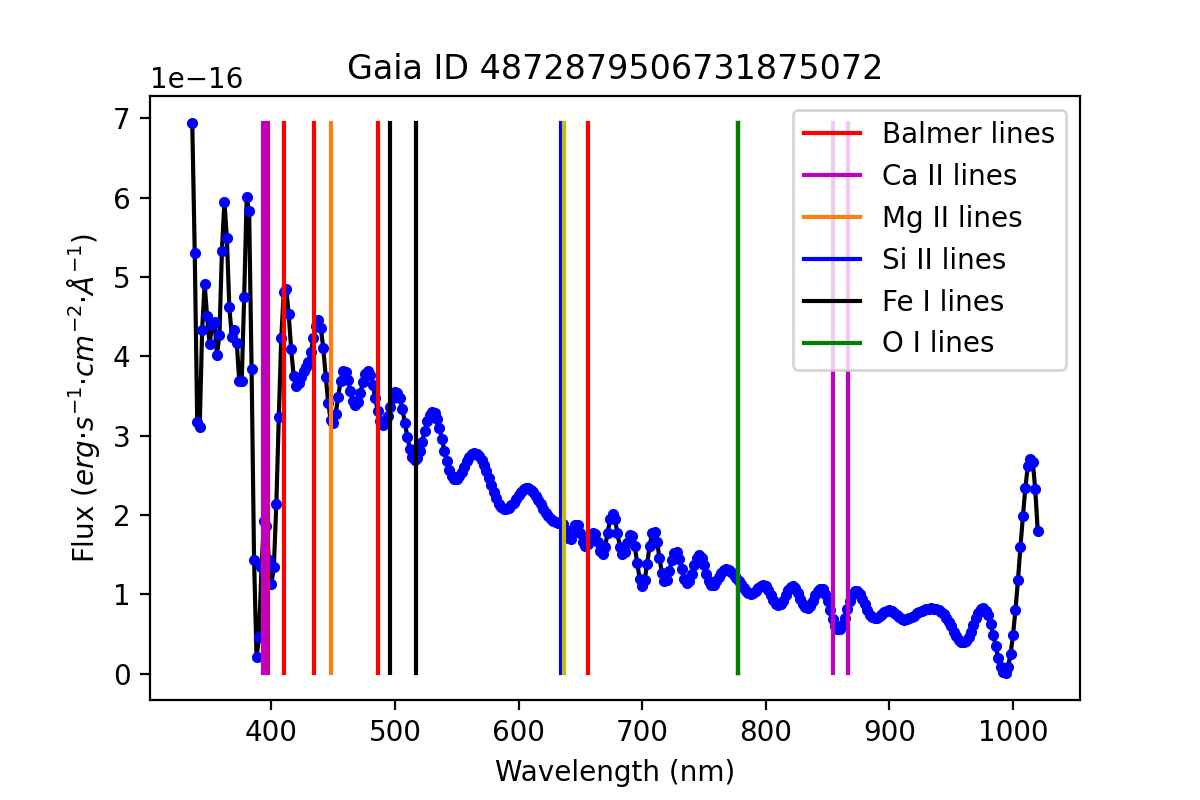}
    \caption{As Fig. \ref{f:DA_MWDD_RF} but for a DZ classified in MWDD (left panel) and  a DZ classified by our algorithm (right panel).}
    \label{f:DZ_MWDD_RF}
        \vspace{0.5cm}
\end{figure*}

\subsection{DZ spectra}
\label{9.6}

Metallic lines are present in DZ spectra. In these example (Figure \ref{f:DZ_MWDD_RF}) we have marked spectral lines of elements that frequently form planetary matter: calcium (\ion{Ca}{II} at 3\,933, 3\,968, 8\,542 and 8\,662 $\AA$), magnesium (\ion{Mg}{II} at 4\,481 $\AA$), oxygen (\ion{O}{I} at 7\,772, 7\,774 and 7\,775 $\AA$), iron (\ion{Fe}{I} at 4\,578, 5\,167, 5\,227 and 5\,269 $\AA$) and silicon (\ion{Si}{II} at 6\,347 and 6\,371 $\AA$).

\end{document}